\documentclass[a4paper,11pt]{article}
\usepackage{jheppub}
\usepackage[T1]{fontenc}
\usepackage{graphicx,subcaption}
\usepackage{braket}
\usepackage{bm}
\usepackage{xcolor}

\usepackage{acro}
\DeclareAcronym{gr}{
    short=GR,
    long=general relativity
}
\DeclareAcronym{uv}{
    short=UV,
    long=ultraviolet
}
\DeclareAcronym{ir}{
    short=IR,
    long=infrared
}
\DeclareAcronym{rg}{
    short=RG,
    long=renormalization group
}
\DeclareAcronym{qg}{
    short=QG,
    long=quantum gravity
}
\DeclareAcronym{qft}{
    short=QFT,
    long=quantum field theory
}
\DeclareAcronym{vev}{
    short=VEV,
    long=vacuum expectation value
}
\DeclareAcronym{flrw}{
    short=FLRW,
    long=Friedmann-Lemaitre-Robertson-Walker
}
\DeclareAcronym{sm}{
    short=SM,
    long=standard model
}
\DeclareAcronym{bch}{
    short=BCH,
    long=Baker-Campbell-Hausdorff 
}
\DeclareAcronym{pgws}{
    short=PGWs ,
    long={Primordial Gravitational Waves}
}
\allowdisplaybreaks[1]

\title{Squeezed Gravitons and One-Loop Self-Energy under Light-Cone Smearing}

\author[a,b,c,d]{Hiroki Matsui}
\affiliation[a]{Osaka Central Advanced Mathematical Institute (OCAMI), Osaka Metropolitan
University, 3-3-138 Sugimoto, Sumiyoshi, Osaka 558-8585, Japan}
\affiliation[b]{Department of Physics, College of Humanities and Sciences, Nihon University, Sakurajosui, Tokyo 156-8550, Japan}
\affiliation[c]{Center for Gravitational Physics and Quantum Information, Yukawa Institute for Theoretical Physics, Kyoto University, Kitashirakawa Oiwakecho, Sakyo-ku, Kyoto 606-8502, Japan }
\affiliation[d]{TRIP Headquarters, RIKEN, Wako 351-0198, Japan}
\emailAdd{hiroki.matsui@yukawa.kyoto-u.ac.jp}
\date{\today}

\abstract{
We investigate light-cone smearing induced by quantum fluctuations of
gravitons and its implications for the ultraviolet structure of quantum field theory. By treating the first-order correction to Synge's world function as an operator, we show that the retarded Green's function is smeared by the variance of graviton fluctuations.  The smearing width depends on the quantum state of gravitons: vacuum fluctuations generate a Gaussian smearing of the light cone, coherent states shift the light-cone position, and squeezed states modify the smearing width itself.  We then apply the smeared Feynman propagator to one-loop self-energies in interacting scalar field theories.  Within a phenomenological finite-resolution prescription that keeps the effective width nonzero, the ultraviolet-sensitive short-distance terms in the $\phi^3$ bubble and $\phi^4$ tadpole diagram expressions are finite. For the illustrative choice that the coarse-graining scale is the Planck length, primordial gravitons induce a relative correction of order $10^{-10}$ to the one-loop self-energy. Our results suggest that the quantum state of gravitons can leave a finite imprint on the causal and short-distance structure of quantum field theory.
}

\begin{document}
\maketitle
\flushbottom

\section{Introduction}
\label{sec:introduction}

The construction of a fully consistent quantum theory of gravity remains one of the central open problems in theoretical physics. Although the non-renormalizability of perturbative quantum gravity prevents a complete \ac{uv} description, a great deal can still be learned by treating the metric perturbation $h_{\mu\nu}$ as a quantum field on a fixed background. Within this effective approach, gravitons are well-defined excitations whose low-energy interactions with matter fields are calculable, and their fluctuations are expected to leave imprints on observables that probe short-distance or high-precision physics~\cite{Dyson:2013hbl,Blencowe:2012mp,DeLorenci:2014vwa,Oniga:2015lro,Bassi:2017szd,Lagouvardos:2020laf,Guerreiro:2019vbq,Parikh:2020nrd,Kanno:2020usf,Parikh:2020kfh,Parikh:2020fhy,Kanno:2021gpt,Tobar:2023ksi,Hsiang:2024qou,Carney:2023nzz,Kanno:2018cuk,Kanno:2019gqw,Kanno:2024gjt,Takeda:2025cye,Dorlis:2025zzz,Kanno:2025how,Kanno:2025fpz}. A natural and conceptually striking consequence of such fluctuations is that the causal structure of spacetime itself becomes a quantum object~\cite{Wheeler:1955zz,Wheeler:1957mu}: the light cone, sharply defined in classical relativity, is smeared by the quantum nature of $h_{\mu\nu}$.

The idea that such smearing might soften the \ac{uv} divergences of \ac{qft} dates back to Landau and Pauli, who observed that these divergences originate in the light-cone singularities of two-point functions and might be removed  by quantum metric fluctuations. This conjecture was elaborated by Refs.~\cite{Deser:1957zz,Isham:1970aw,Isham:1972pf,
Ford:1994cr,Ford:1996qc,Ohanian:1996ni,Ohanian:1999fu,
Modesto:2009qc,Abel:2019ufz,Padmanabhan:2020rba,Kan:2020vut}. The natural geometric object encoding this smearing effect is Synge's world function $\sigma (x,x')$, defined as one half of the squared geodesic interval. Promoting the linear shift $\sigma (x,x')$ to an operator $\hat{\sigma}(x,x')$ on the Hilbert space allows the retarded Green's function to be written as a state-dependent average, whose evaluation translates the classical singularity into a smeared Gaussian distribution. Although gravity-induced minimum lengths have been discussed~\cite{Garay:1994en}, the present construction only smears the light-cone singularity for separated points. It does not derive a fundamental minimal length or remove the coincident-point singularity within the leading-order continuum treatment.

Despite the conceptual clarity of this framework, the detailed dependence of the light-cone smearing on the specific quantum state of the graviton field has remained largely unexplored. This is a significant gap, because the magnitude and structure of the smearing are determined by the variance $\langle \hat{\sigma}^2 (x,x')\rangle$, which is an intrinsically state-dependent quantity~\cite{Ford:1994cr}. Two physically distinct situations are particularly relevant. First, classical gravitational waves are most naturally described by coherent states, which saturate the minimum-uncertainty bound and approximate classical field configurations. Second, primordial gravitons generated during inflation evolve into highly squeezed two-mode states with extremely large squeezing parameters at late times~\cite{Grishchuk:1990bj,Albrecht:1992kf,Polarski:1995jg}. The squeezed character of these primordial modes is a robust prediction of standard inflationary cosmology and underlies many discussions of the quantum-to-classical transition of cosmological perturbations. It is therefore of basic interest to ask how each of these distinct quantum states contributes to the smearing of the light cone, and whether classical and genuinely non-classical gravitational fields leave qualitatively different imprints on field propagation.

A second motivation comes from the structure of perturbative \ac{qft}. Loop amplitudes in interacting theories are plagued by short-distance singularities, traceable in coordinate space to the coincident-point limit of the Feynman propagator. Standard regularization schemes such as dimensional regularization or momentum cutoffs are technically convenient but lack a direct physical interpretation. If quantum gravitational fluctuations smear the light cone, one may ask whether the corresponding propagator becomes regular at coincidence and provides a state-dependent regularization of \ac{uv} divergences. This possibility, however, was not realized in the original analysis using $\langle \hat{\sigma}^2 (x,x')\rangle$: the one-loop electron self-energy computed there retained residual coincident-point singularities~\cite{Ford:1994cr}, and the search for a one-loop process actually rendered finite by metric fluctuations was explicitly identified as a question.

In this paper, we address both issues. Following the approach of Ref.~\cite{Ford:1994cr}, we develop the operator-level treatment of light-cone smearing for several physically motivated graviton states and apply the resulting smeared propagator to one-loop self-energies in scalar field theory. We compute the expectation value of the retarded Green's function in the graviton vacuum, in coherent states, and in squeezed vacuum states, and show that each yields a Gaussian smearing of the classical light cone whose width is controlled by the state-dependent variance $\langle \hat{\sigma}^2 (x,x')\rangle$. Coherent states are found to merely shift the location of the light cone, in agreement with the classical limit, while squeezed states genuinely modify the smearing width and can either enhance or suppress it depending on the squeezed quadrature.

We then use the smeared Feynman propagator to evaluate the bubble self-energy in $\phi^3$ theory and the tadpole self-energy in $\phi^4$ theory. Within a coarse-grained model, a prescribed nonzero smearing width acts as an effective regulator while reproducing the expected long-distance logarithmic structure. In particular, the coincident-point singularity of the $\phi^4$ tadpole is replaced by a finite expression at fixed nonzero width. A subtle point, however, is that the leading correction $\hat{\sigma}_1$ induced by the metric perturbation is proportional to the coordinate separation between the two points. Consequently, in the strict coincidence limit $x'\to x$, the leading-order point-split variance $\langle \hat{\sigma}_1^2(x,x')\rangle$ formally vanishes. The perturbative continuum formalism at the leading-order metric perturbation therefore neither derives a fundamental minimal length nor removes the coincident-point singularity by itself. We instead introduce a finite coarse-graining scale as a phenomenological assumption in order to explore the consequences of limited spacetime resolution.
Finally, by incorporating the contribution of long-wavelength primordial gravitons generated during inflation, we estimate a relative correction to the one-loop self-energy, and discuss the prospects for testing such effects in precision measurements like the lepton anomalous magnetic moment.

The remainder of this paper is organized as follows. In Sec.~\ref{sec:retarded-Greens-function}, we review the formalism of light-cone fluctuations based on Synge's world function and the retarded Green's function. In Sec.~\ref{sec:quantum-state-graviton}, we evaluate the smearing for the graviton vacuum, coherent, and two-mode squeezed states, and clarify the role of the underlying quantum state. In Sec.~\ref{sec:one-loop-self-energy}, we apply the smeared propagator to the one-loop bubble and tadpole self-energies and analyze the implications of the inflationary graviton contribution. We summarize our results and discuss future directions in Sec.~\ref{sec:conclusion}. Several technical details are relegated to the appendices. Throughout this paper, we adopt the metric signature $(-,+,+,+)$ and use natural units $c=\hbar=1$, with the reduced Planck mass denoted by $M_{\rm pl}$.

\section{Retarded Green's function and Synge's world function}
\label{sec:retarded-Greens-function}

In this section, we review the formalism of light-cone fluctuations induced by graviton fluctuations around a flat background, following the approach developed
in Ref.~\cite{Ford:1994cr}. 
The basic geometrical object in this formalism is
Synge's world function $\sigma(x,x')$, which is defined as one half of the squared geodesic interval between two spacetime points $x$ and $x'$~\cite{synge1960special}. Synge's world function provides a covariant way to characterize the
local geometry of spacetime. In particular, in a geodesic neighborhood, the metric can be recovered from the coincidence limit of its mixed second derivative.  With our sign convention, this relation is
\begin{equation}
g_{\mu\nu}(x)
=
-\lim_{x'\to x}
\frac{\partial}{\partial x^\mu}
\frac{\partial}{\partial x'^\nu}
\sigma(x,x') .
\label{eq:metric_from_synge}
\end{equation}

In a weakly perturbed spacetime, the metric is decomposed into the Minkowski background and a small perturbation $h_{\mu\nu}$ representing the graviton field,
\begin{equation}\label{eq:metric-perturbations}
g_{\mu\nu} = \eta_{\mu\nu} + h_{\mu\nu}, \qquad |h_{\mu\nu}| \ll 1.
\end{equation}
Correspondingly, the world function admits the expansion
\begin{equation}\label{eq:Sw-function-metric-perturbations}
\sigma(x,x') = \sigma_0(x,x') + \sigma_1(x,x') + \mathcal{O}(h^2),
\end{equation}
where $\sigma_0$ denotes the world function in Minkowski spacetime,
\begin{equation}
\sigma_0(x, x') = \frac{1}{2} \eta_{\mu\nu}\,z^\mu z^\nu 
= \frac{1}{2}(x-x')^2 
= \frac{1}{2}\!\left[ -(t-t')^2 + |\mathbf{x}-\mathbf{x}'|^2 \right],
\end{equation}
and we have introduced the coordinate separation $z^\mu \equiv x^\mu - x'^\mu$. The first-order correction $\sigma_1$ encodes the shift of the geodesic interval induced by $h_{\mu\nu}$, and is obtained by integrating the perturbation along the unperturbed geodesic,
\begin{equation}\label{eq:sigma1-classical}
\sigma_1(x,x')
= \frac{1}{2}\int_{0}^{1} d\lambda\;
h_{\mu\nu}\!\left(\bar x(\lambda)\right)\,z^\mu z^\nu,
\end{equation}
where $\bar x(\lambda) = x' + \lambda z$ parameterizes the background geodesic from $x'$ to $x$ with affine parameter $\lambda \in [0, 1]$. When $h_{\mu\nu}$ varies slowly over the scale of the separation, this reduces to $\sigma_1 \simeq \frac{1}{2} h_{\mu\nu} z^\mu z^\nu$. In the following analysis, we work to leading order in $h_{\mu\nu}$ and neglect contributions of $\mathcal{O}(h^2)$ and higher.

To investigate the influence of graviton fluctuations on field propagation, we consider a massless scalar field $\hat{\phi}(x)$ on the perturbed background. In the framework of \ac{qft}, the causal propagation of field disturbances is encoded in the retarded Green's function $G_{\rm ret}(x,x')$. In flat spacetime, this is the causal fundamental solution to the wave equation,
\begin{equation}\label{eq:Gret-flat}
G_{\rm ret}^{(0)}(x,x') = \frac{\theta(t-t')}{4\pi}\,\delta(\sigma_0),
\end{equation}
which is supported on the future light cone of the source point. The detailed derivation from the field commutator is provided in Appendix~\ref{app:QFT-basics}.

In the perturbed spacetime, the light-cone structure is modified by $h_{\mu\nu}$. To leading order, the retarded Green's function is obtained by replacing $\sigma_0$ in Eq.~\eqref{eq:Gret-flat} with the full world function $\sigma = \sigma_0 + \sigma_1$, yielding
\begin{equation}
G_{\rm ret}(x,x') \simeq \frac{\theta(t-t')}{4\pi}\,\delta\!\left(\sigma_0 + \sigma_1\right).
\end{equation}
Using the Fourier representation of the delta function, this can be rewritten as
\begin{equation}\label{eq:Gret-fourier}
G_{\rm ret}(x,x')
\simeq \frac{\theta(t-t')}{8\pi^2}\int_{-\infty}^{\infty}du\,e^{iu\sigma_0}\,e^{iu\sigma_1}.
\end{equation}

While $h_{\mu\nu}$ and $\sigma_1$ have been treated as classical quantities thus far, a consistent treatment of quantum gravity requires that the metric perturbations be promoted to operators acting on the graviton Hilbert space. The first-order shift of Synge's world function then becomes the operator
\begin{equation}\label{eq:sigma1-operator}
\hat{\sigma}_1(x,x')
= \frac{1}{2}\int_{0}^{1} d\lambda\,
\hat{h}_{\mu\nu}\!\left(\bar{x}(\lambda)\right)\,
z^\mu z^\nu.
\end{equation}
For a given graviton state $\ket{\psi}$, the metric-averaged retarded Green's function is determined by the expectation value of the exponential operator,
\begin{equation}\label{eq:Gret-general-state}
\langle G_{\rm ret}(x,x')\rangle
\simeq
\frac{\theta(t-t')}{8\pi^2}
\int_{-\infty}^{\infty}du\,
e^{iu\sigma_0}\,
\bra{\psi} e^{iu\hat{\sigma}_1} \ket{\psi}.
\end{equation}
While the formalism of light-cone fluctuations was introduced in Ref.~\cite{Ford:1994cr}, a systematic analysis of how the smearing depends on the specific quantum state of the graviton field has remained unexplored. In the next section, we therefore investigate how the retarded Green's function and the observable field propagation are modified for several graviton states.

\section{Quantum states of gravitons and light-cone smearing}
\label{sec:quantum-state-graviton}

In this section, we evaluate the expectation value of the retarded Green's function $\langle G_{\rm ret}\rangle = \bra{\psi}\hat{G}_{\rm ret}\ket{\psi}$ for several graviton states, beginning with the vacuum states and then turning to coherent and two-mode squeezed vacuum states.

\subsection{Vacuum state}
\label{subsec:vacuum-state}

The operator $\hat{\sigma}_1$ defined in Eq.~\eqref{eq:sigma1-operator} is linear in the graviton field $\hat{h}_{\mu\nu}$, so it admits a standard decomposition into positive and negative frequency parts,
$\hat{\sigma}_1 = \hat{\sigma}_1^+ + \hat{\sigma}_1^-$, defined by
\begin{equation}
\hat{\sigma}_1^+\ket{0}=0,
\quad
\bra{0}\hat{\sigma}_1^-=0.
\end{equation}
In terms of the annihilation and creation operators of the mode basis defining the vacuum $\ket{0}$, we write
\footnote{Equivalently, by relabeling $\bm k\to-\bm k$, one has
\begin{equation*}
\hat{\sigma}_1^-
=\sum_{\bm k}
\hat{a}_{\bm k}^{\dagger}\,f_k^*\,e^{-i\bm k\cdot\bm x},
\end{equation*}
which manifestly satisfies the Hermiticity relation
$\hat{\sigma}_1^-=(\hat{\sigma}_1^+)^\dagger$.}
\begin{equation}\label{eq:sigma1-pm}
\hat{\sigma}_1^+ = \sum_{\bm k}\hat{a}_{\bm k}\, f_k\, e^{i\bm k\cdot\bm x},
\qquad
\hat{\sigma}_1^- = \sum_{\bm k}\hat{a}_{-\bm k}^{\dagger}\, f_k^*\, e^{i\bm k\cdot\bm x},
\end{equation}
where $f_k$ are the mode functions.

To evaluate the expectation value of the exponential operator $e^{iu\hat{\sigma}_1}$, we use the \ac{bch} formula. Since $\hat{\sigma}_1$ is linear in the field operators, the commutator $[\hat{\sigma}_1^+,\hat{\sigma}_1^-]$ is a $c$-number. Using the canonical commutation relation $[\hat{a}_{\bm k},\hat{a}_{-\bm l}^{\dagger}]=\delta_{\bm k,-\bm l}$ and the isotropy condition $f_k=f_{-k}$, we find
\begin{equation}
[\hat{\sigma}_1^+,\hat{\sigma}_1^-]
=
\sum_{\bm k,\bm l}
f_k\,f_l^*\, e^{i(\bm k+\bm l)\cdot\bm x}
[\hat{a}_{\bm k},\hat{a}_{-\bm l}^{\dagger}]
=\sum_{\bm k}|f_k|^2,
\end{equation}
which coincides with the vacuum variance of the world function operator,
\begin{equation}
\langle\hat{\sigma}_1^2\rangle
\equiv \bra{0}\hat{\sigma}_1^2\ket{0}
=\bra{0}[\hat{\sigma}_1^+, \hat{\sigma}_1^-]\ket{0}
=\sum_{\bm k}|f_k|^2.
\end{equation}

Applying the \ac{bch} identity $e^{\hat{A}+\hat{B}} = e^{\hat{B}} e^{\hat{A}} e^{\frac{1}{2}[\hat{A}, \hat{B}]}$,
\footnote{If $\hat{A}$ and $\hat{B}$ are noncommuting operators whose commutator commutes with both of them, $[\hat{A},[\hat{A},\hat{B}]]=[\hat{B},[\hat{A},\hat{B}]]=0$, the \ac{bch} formula yields $e^{\hat{A}+\hat{B}} = e^{\hat{B}} e^{\frac{1}{2}[\hat{A}, \hat{B}]} e^{\hat{A}}$.}
the phase-factor operator can be normal-ordered as
\begin{equation}\label{eq:exp-normal-ordered}
e^{iu\hat{\sigma}_1} = e^{iu\hat{\sigma}_1^-}\, e^{iu\hat{\sigma}_1^+}\, e^{-\frac{1}{2}u^2 [\hat{\sigma}_1^+, \hat{\sigma}_1^-]}.
\end{equation}
For the graviton vacuum, $e^{iu\hat{\sigma}_1^+}\ket{0} = \ket{0}$ and $\bra{0}e^{iu\hat{\sigma}_1^-} = \bra{0}$, so the expectation value reduces to
\begin{equation}\label{eq:exp-sigma1-vacuum}
\langle e^{iu\hat{\sigma}_1}\rangle
= \bra{0} e^{iu\hat{\sigma}_1^-} e^{iu\hat{\sigma}_1^+} \ket{0}\,
e^{-\frac{1}{2}u^2\langle \hat{\sigma}_1^2 \rangle}
= e^{-\frac{1}{2}u^2\langle \hat{\sigma}_1^2 \rangle},
\end{equation}
which exhibits Gaussian decay in $u$ controlled by the vacuum variance $\langle \hat{\sigma}_1^2 \rangle$.

Substituting Eq.~\eqref{eq:exp-sigma1-vacuum} into Eq.~\eqref{eq:Gret-general-state}, the expectation value of the  retarded Green's function becomes
\begin{equation}
\langle G_{\rm ret}(x,x') \rangle
= \frac{\theta(t-t')}{8\pi^2}\int_{-\infty}^{\infty}du\,e^{iu\sigma_0}\,
e^{-\frac{1}{2}u^2\langle \hat{\sigma}_1^2 \rangle},
\end{equation}
which converges for $\langle \hat{\sigma}_1^2 \rangle>0$. Performing the Gaussian integral over $u$, we obtain
\footnote{Including the second-order correction $\hat{\sigma}_2=\mathcal{O}(h^2)$ to Synge's world function, $\langle\hat{\sigma}_2\rangle$ merely shifts the peak location of the expectation value of the retarded Green's function, while the smearing width is governed by $\langle\hat{\sigma}_1^2\rangle$~\cite{Ford:1994cr}. We therefore neglect $\hat{\sigma}_2$ throughout and focus on the leading smearing controlled by $\hat{\sigma}_1$.}
\begin{equation}\label{eq:Gret-vacuum}
\langle G_{\rm ret}(x,x') \rangle
=\frac{\theta(t-t')}{8\pi^2}
\sqrt{\frac{2\pi}{\langle \hat{\sigma}_1^2 \rangle}}\,
\exp\!\left[-\frac{\sigma_0^2}{2\langle \hat{\sigma}_1^2 \rangle}\right].
\end{equation}
The classical light-cone singularity $\delta(\sigma_0)$ is thereby smeared into a Gaussian distribution whose width is set by the graviton vacuum fluctuations. In the limit $\langle \hat{\sigma}_1^2 \rangle \to 0$, the standard distributional form is recovered,
\begin{equation}
\langle G_{\rm ret}(x,x') \rangle
\;\longrightarrow\;
\frac{\theta(t-t')}{4\pi}\,\delta(\sigma_0).
\end{equation}
Since the smearing depends on the underlying quantum state of the gravitational field, we now generalize the vacuum result to coherent and squeezed vacuum states, both of which are of particular interest in cosmological and gravitational wave contexts.

\subsection{Coherent state}
\label{subsec:coherent-state}

We now consider the coherent state $\ket{\xi}\equiv\hat{D}(\xi)\ket{0}$, generated by acting on the vacuum with the displacement operator
\begin{equation}
\hat{D}(\xi)=\exp\!\left[\sum_{\bm k}\left(\xi_{\bm k}\,\hat{a}_{\bm k}^\dagger - \xi_{\bm k}^*\,\hat{a}_{\bm k}\right)\right],
\end{equation}
where $\xi_{\bm k}$ is the complex displacement parameter for mode $\bm k$. The coherent state is an eigenstate of the annihilation operator,
\begin{equation}
\hat{a}_{\bm k}\ket{\xi}=\xi_{\bm k}\ket{\xi},
\end{equation}
and represents a minimum-uncertainty state corresponding to a classical field configuration.

Acting with $\hat{\sigma}_1^{+}$ on the coherent state yields
\begin{align}
\hat{\sigma}_1^+\ket{\xi}
&=\sum_{\bm k} f_k\,e^{i\bm k\cdot\bm x}\,\hat{a}_{\bm k}\ket{\xi}
\nonumber\\
&=\left(\sum_{\bm k} f_k\,e^{i\bm k\cdot\bm x}\,\xi_{\bm k}\right)\ket{\xi}
\equiv\Xi\ket{\xi},
\end{align}
and similarly $\bra{\xi}\hat{\sigma}_1^{-}=\bra{\xi}\Xi^*$. Combining this with the normal-ordered form Eq.~\eqref{eq:exp-normal-ordered}, we obtain
\begin{align}
\langle e^{iu\hat{\sigma}_1}\rangle_{\rm coh}
&\equiv \bra{\xi}e^{iu\hat{\sigma}_1}\ket{\xi}
\notag\\
&=\exp\!\left(-\tfrac{1}{2}u^2\langle\hat{\sigma}_1^2\rangle\right)
\bra{\xi}e^{iu\hat{\sigma}_1^{-}}e^{iu\hat{\sigma}_1^{+}}\ket{\xi}
\notag\\
&=\exp\!\left[-\tfrac{1}{2}u^2\langle\hat{\sigma}_1^2\rangle + iu\,(\Xi+\Xi^*)\right]
\notag\\
&=\exp\!\left[-\tfrac{1}{2}u^2\langle\hat{\sigma}_1^2\rangle + iu\langle\hat{\sigma}_1\rangle_{\rm coh}\right],
\end{align}
where the mean value is shifted by $\langle\hat{\sigma}_1\rangle_{\rm coh} = \Xi+\Xi^* = 2\,\mathrm{Re}\,\Xi$, while the Gaussian width is unchanged from the vacuum case.

After averaging over the coherent-state graviton fluctuations, the retarded Green's function becomes
\begin{equation}
\langle G_{\rm ret}(x,x') \rangle_{\rm coh}
=\frac{\theta(t-t')}{8\pi^2}
\int_{-\infty}^{\infty}du\,
\exp\!\left[iu\bigl(\sigma_0+\langle\hat{\sigma}_1\rangle_{\rm coh}\bigr)\right]
e^{-\frac{1}{2}u^2\langle\hat{\sigma}_1^2\rangle},
\end{equation}
and evaluating the Gaussian integral gives
\begin{equation}\label{eq:Gret-coherent}
\langle G_{\rm ret}(x,x') \rangle_{\rm coh}
=\frac{\theta(t-t')}{8\pi^2}
\sqrt{\frac{2\pi}{\langle\hat{\sigma}_1^2\rangle}}\,
\exp\!\left[-\frac{\bigl(\sigma_0+\langle\hat{\sigma}_1\rangle_{\rm coh}\bigr)^2}{2\langle\hat{\sigma}_1^2\rangle}\right].
\end{equation}
This expression is identical to the vacuum result Eq.~\eqref{eq:Gret-vacuum} after the substitution $\sigma_0 \to \sigma_0+\langle\hat{\sigma}_1\rangle_{\rm coh}$. Since the coherent state corresponds to a classical configuration of the gravitational field, this confirms that a classical gravitational wave merely displaces the location of the light cone and does not contribute to its quantum smearing.

The same conclusion follows from a complementary viewpoint, namely splitting the metric into a classical wave plus quantum vacuum fluctuations, $h_{\mu\nu}=h^{\rm cl}_{\mu\nu}+\delta\hat{h}_{\mu\nu}$. The world-function operator then decomposes as
\begin{equation}
\hat{\sigma}_1 = \sigma_1^{\rm cl} + \delta\hat{\sigma}_1,
\end{equation}
where $\sigma_1^{\rm cl}$ is the $c$-number classical contribution and $\delta\hat{\sigma}_1$ is the operator-valued vacuum fluctuation. Since $\sigma_1^{\rm cl}$ commutes with $\delta\hat{\sigma}_1$, the vacuum expectation value factorizes,
\begin{equation}
\langle e^{iu\hat{\sigma}_1}\rangle
= \bra{0}e^{iu(\sigma_1^{\rm cl}+\delta\hat{\sigma}_1)}\ket{0}
= e^{iu\sigma_1^{\rm cl}}\,\bra{0}e^{iu\delta\hat{\sigma}_1}\ket{0}
= e^{iu\sigma_1^{\rm cl}}\,e^{-\frac{1}{2}u^2\langle\delta\hat{\sigma}_1^2\rangle},
\end{equation}
and the resulting Green's function is
\begin{equation}
\langle G_{\rm ret}(x,x') \rangle
=\frac{\theta(t-t')}{8\pi^2}
\sqrt{\frac{2\pi}{\langle\delta\hat{\sigma}_1^2\rangle}}\,
\exp\!\left[-\frac{(\sigma_0+\sigma_1^{\rm cl})^2}{2\langle\delta\hat{\sigma}_1^2\rangle}\right].
\end{equation}
Hence the light-cone smearing is governed solely by the quantum fluctuations $\delta\hat{\sigma}_1$, while the classical gravitational wave contributes only a shift of the light cone. Genuine smearing therefore requires a non-classical graviton state, and we now turn to the squeezed vacuum and show that its effect on the light cone is qualitatively different from that of a coherent state.

\subsection{Squeezed vacuum state}
\label{subsec:squeezed-state}

We finally consider the squeezed vacuum state $\ket{\zeta}\equiv\hat{S}(\zeta)\ket{0}$. In standard inflationary cosmology, primordial gravitons originating from the Bunch-Davies vacuum are naturally described at late times by two-mode squeezed states involving the pair of modes $\bm k$ and $-\bm k$. We therefore adopt the two-mode squeezing operator
\begin{equation}
\hat{S}(\zeta)
=\exp\!\left[
\sum_{\bm k}\left(\zeta_k^*\,\hat{a}_{\bm k}\hat{a}_{-\bm k} - \zeta_k\,\hat{a}_{\bm k}^\dagger\hat{a}_{-\bm k}^\dagger\right)\right],
\end{equation}
where $\zeta_k = r_k\,e^{i\theta_k}$ is the squeezing parameter with amplitude $r_k\geq0$ and phase $\theta_k$, and the summation is restricted to a momentum half-space to avoid double-counting the pair $(\bm k,-\bm k)$. Spatial isotropy implies $\zeta_{-k}=\zeta_k$.

Unlike the vacuum and coherent states, the squeezed vacuum is not an eigenstate of $\hat{a}_{\bm k}$ and hence not of $\hat{\sigma}_1^+$. It is therefore more convenient to evaluate the expectation value by using the unitary transformation generated by the squeezing operator,
\footnote{Since $\hat{S}(\zeta)$ is unitary, $\hat{S}^\dagger(\zeta)=\hat{S}^{-1}(\zeta)$, and for any operator $\hat{A}$ one has $\hat{S}^\dagger(\zeta)e^{\hat{A}}\hat{S}(\zeta) = e^{\hat{S}^\dagger(\zeta)\hat{A}\hat{S}(\zeta)}$.}
\begin{equation}
\langle e^{iu\hat{\sigma}_1}\rangle_{\rm sq}
\equiv \bra{\zeta}e^{iu\hat{\sigma}_1}\ket{\zeta}
= \bra{0}\hat{S}^\dagger(\zeta)e^{iu\hat{\sigma}_1}\hat{S}(\zeta)\ket{0}
= \bra{0}e^{iu\hat{\Sigma}_1}\ket{0},
\end{equation}
where we have introduced the transformed operator
\begin{equation}
\hat{\Sigma}_1
\equiv \hat{S}^\dagger(\zeta)\,\hat{\sigma}_1\,\hat{S}(\zeta).
\end{equation}

The Bogoliubov transformation of the annihilation and creation operators reads
\begin{align}
\hat{b}_{\bm k}
&\equiv \hat{S}^\dagger(\zeta)\hat{a}_{\bm k}\hat{S}(\zeta)
= \hat{a}_{\bm k}\cosh r_k - \hat{a}_{-\bm k}^{\dagger}\,e^{i\theta_k}\sinh r_k,
\\
\hat{b}_{-\bm k}^{\dagger}
&\equiv \hat{S}^\dagger(\zeta)\hat{a}_{-\bm k}^{\dagger}\hat{S}(\zeta)
= \hat{a}_{-\bm k}^{\dagger}\cosh r_k - \hat{a}_{\bm k}\,e^{-i\theta_k}\sinh r_k,
\end{align}
with the corresponding Bogoliubov coefficients
\begin{equation}
\alpha_k=\cosh r_k,\quad \beta_k=e^{-i\theta_k}\sinh r_k.
\end{equation}
Substituting these into
\begin{equation}
\hat{\sigma}_1=
\sum_{\bm k}\left(f_k\,\hat{a}_{\bm k} + f_k^*\,\hat{a}_{-\bm k}^{\dagger}\right)e^{i\bm k\cdot\bm x},
\end{equation}
we obtain
\begin{align}
\hat{\Sigma}_1
&=\sum_{\bm k}\Big[
f_k\bigl(\hat{a}_{\bm k}\cosh r_k - \hat{a}_{-\bm k}^{\dagger}\,e^{i\theta_k}\sinh r_k\bigr)
\nonumber\\
&\hspace{3.5em}
+ f_k^*\bigl(\hat{a}_{-\bm k}^{\dagger}\cosh r_k - \hat{a}_{\bm k}\,e^{-i\theta_k}\sinh r_k\bigr)
\Big]e^{i\bm k\cdot\bm x}
\nonumber\\
&=\sum_{\bm k}\bigl[g_k\,\hat{a}_{\bm k} + g_k^*\,\hat{a}_{-\bm k}^{\dagger}\bigr]e^{i\bm k\cdot\bm x},
\end{align}
where the squeezed mode coefficient is
\begin{equation}\label{eq:gk-def}
g_k \equiv f_k\cosh r_k - f_k^*\,e^{-i\theta_k}\sinh r_k.
\end{equation}

The transformed operator $\hat{\Sigma}_1$ has the same algebraic structure as $\hat{\sigma}_1$, with $f_k$ replaced by $g_k$. Decomposing $\hat{\Sigma}_1=\hat{\Sigma}_1^+ + \hat{\Sigma}_1^-$ and applying the same \ac{bch} argument as in the vacuum case, we obtain
\begin{equation}
\bra{0}e^{iu\hat{\Sigma}_1}\ket{0}
=e^{-\frac{1}{2}u^2\langle\hat{\Sigma}_1^2\rangle},
\quad
\langle\hat{\Sigma}_1^2\rangle
=\bra{0}[\hat{\Sigma}_1^+,\hat{\Sigma}_1^-]\ket{0}
=\sum_{\bm k}|g_k|^2.
\end{equation}
Since $\langle e^{iu\hat{\sigma}_1}\rangle_{\rm sq}=\bra{0}e^{iu\hat{\Sigma}_1}\ket{0}$, this quantity is precisely the variance of $\hat{\sigma}_1$ in the squeezed vacuum,
\begin{equation}
\langle\hat{\sigma}_1^2\rangle_{\rm sq}
=\langle\hat{\Sigma}_1^2\rangle
=\sum_{\bm k}|g_k|^2.
\end{equation}

An explicit evaluation of $|g_k|^2$ from Eq.~\eqref{eq:gk-def} gives
\begin{align}
|g_k|^2 
&= \bigl(f_k\cosh r_k - f_k^*e^{-i\theta_k}\sinh r_k\bigr)\bigl(f_k^*\cosh r_k - f_k\,e^{i\theta_k}\sinh r_k\bigr)
\notag\\
&= |f_k|^2\cosh(2r_k) - \mathrm{Re}\bigl(f_k^2\,e^{i\theta_k}\bigr)\sinh(2r_k),
\label{eq:gk-squared}
\end{align}
To make the meaning of squeezing and anti-squeezing explicit, let
$f_k=|f_k|e^{i\varphi_k}$. Equation~\eqref{eq:gk-squared} then becomes
\begin{equation*}
|g_k|^2
=|f_k|^2\left[
\cosh(2r_k)
-\cos(2\varphi_k+\theta_k)\sinh(2r_k)
\right],
\end{equation*}
and consequently
\begin{equation*}
e^{-2r_k}|f_k|^2
\leq |g_k|^2
\leq e^{2r_k}|f_k|^2.
\end{equation*}
Here, by squeezing we mean a phase choice for which
$|g_k|^2<|f_k|^2$, so that the contribution of mode $k$ to the
world-function variance is suppressed relative to its vacuum value.
By anti-squeezing we mean $|g_k|^2>|f_k|^2$, for which that
contribution is enhanced. In particular,
$2\varphi_k+\theta_k=0$ gives
$|g_k|^2=e^{-2r_k}|f_k|^2$, whereas
$2\varphi_k+\theta_k=\pi$ gives
$|g_k|^2=e^{2r_k}|f_k|^2$.
so that the expectation value in the squeezed vacuum is
\begin{equation}
\langle e^{iu\hat{\sigma}_1}\rangle_{\rm sq}
= \exp\!\left[-\tfrac{1}{2}u^2\sum_{\bm k}\!\left(|f_k|^2\cosh(2r_k) - \mathrm{Re}\bigl(f_k^2\,e^{i\theta_k}\bigr)\sinh(2r_k)\right)\right],
\end{equation}
and the corresponding variance reads
\begin{equation}\label{eq:variance-squeezed}
\langle\hat{\sigma}_1^2\rangle_{\rm sq}
= \sum_{\bm k}\!\left[|f_k|^2\cosh(2r_k) - \mathrm{Re}\bigl(f_k^2\,e^{i\theta_k}\bigr)\sinh(2r_k)\right].
\end{equation}

Consequently, the averaged retarded Green's function in the squeezed vacuum is
\begin{equation}\label{eq:Gret-squeezed}
\langle G_{\rm ret}(x,x') \rangle_{\rm sq}
= \frac{\theta(t-t')}{8\pi^2}
\sqrt{\frac{2\pi}{\langle\hat{\sigma}_1^2\rangle_{\rm sq}}}\,
\exp\!\left[-\frac{\sigma_0^2}{2\langle\hat{\sigma}_1^2\rangle_{\rm sq}}\right].
\end{equation}
The smearing of the light-cone singularity is now governed by the squeezed-state variance $\langle\hat{\sigma}_1^2\rangle_{\rm sq}$. In contrast to the coherent state, the squeezed vacuum genuinely modifies the width of the smeared Gaussian, producing a qualitatively non-classical effect. Such squeezed gravitons arise naturally in inflationary cosmology, where the corresponding variance is evaluated explicitly in Appendix~\ref{app:squeezed-gravitons-cosmology}.

The same averaging procedure applies to the Feynman propagator $G_F$, which is the relevant object for perturbative loop computations. As shown in Appendix~\ref{app:smeared-feynman-propagator}, both $\langle G_F(x,x')\rangle$ and $\langle G_F(x,x')^2\rangle$ take Gaussian forms controlled by the state-dependent variance $\langle\hat{\sigma}_1^2\rangle$. For a prescribed nonzero variance, both quantities are finite on the classical light cone. In the following section, we study the corresponding one-loop amplitudes, focusing on the bubble diagram in $\phi^3$ theory and the tadpole diagram in $\phi^4$ theory.

\section{One-loop self-energy under light-cone smearing}
\label{sec:one-loop-self-energy}

In this section, we apply the smeared Feynman propagator derived in Appendix~\ref{app:smeared-feynman-propagator} to one-loop self-energies in interacting scalar field theory. We first analyze the ordinary \ac{qft} result in coordinate space, where the \ac{uv} sensitivity of loop amplitudes can be traced directly to the short-distance singularities of the local Feynman propagator. We then replace the local propagator with its smeared counterpart and examine how the bubble diagram in $\phi^3$ theory and the tadpole diagram in $\phi^4$ theory are modified.

In the loop calculations below, we treat the smearing width as a prescribed
effective coarse-grained parameter,
\begin{equation*}
\Delta_{\sigma,\mathrm{eff}}^2>0,
\end{equation*}
which is assumed to be approximately independent of the point separation
over the range relevant to the integrals. This parameter should not be
confused with the point-split variance
$\langle\hat{\sigma}_1^2(x,x')\rangle$ obtained directly from the
leading-order continuum theory. The latter is generally
separation-dependent and vanishes in the strict coincidence limit, as
discussed below. The following calculations therefore describe a
finite-resolution effective model rather than a first-principles
ultraviolet completion.

Hereafter, we use the Fourier convention of Feynman propagator
\begin{equation}
G_F(z)
=\frac{1}{(2\pi)^4}\int d^4k\,
\frac{e^{-ik\cdot z}}{k^2-m^2+i\epsilon}\,,
\quad
z^\mu \equiv x^\mu-x'^\mu ,
\label{eq:GF_fourier}
\end{equation}
so that $G_F(x,x')\equiv G_F(z)$.  With this convention, the inverse relation is
\begin{equation}
\frac{1}{k^2-m^2+i\epsilon}
=\int d^4z\,e^{ik\cdot z}\,G_F(z).
\label{eq:inverse_transforms}
\end{equation}
Indeed, substituting Eq.~\eqref{eq:GF_fourier} into the right-hand side gives
\begin{align}
\int d^4z\,e^{ik\cdot z}G_F(z)
&=
\frac{1}{(2\pi)^4}\int d^4q\,
\frac{1}{q^2-m^2+i\epsilon}
\int d^4z\,e^{i(k-q)\cdot z}  \notag\\
&=\frac{1}{k^2-m^2+i\epsilon}.
\end{align}

\subsection{$\phi^3$ theory: bubble self-energy and short-distance singularity}
\label{subsubsec:phi3_bubble}

We now consider a real scalar field with cubic interaction,
\begin{equation}
\mathcal{L}_{\rm int}=-\frac{g}{3!}\phi^3 .
\end{equation}
The one-loop self-energy is the bubble diagram,
\begin{align}
\Sigma(p)
&=\frac{i g^2}{2}\int\frac{d^4k}{(2\pi)^4}\,
\frac{1}{k^2-m^2+i\epsilon}\,
\frac{1}{(p-k)^2-m^2+i\epsilon}\,,
\label{eq:Sigma_phi3_momentum}
\end{align}
where the factor $\frac{1}{2}$ is the symmetry factor of the bubble diagram.  Using Eq.~\eqref{eq:inverse_transforms} for both internal propagators, we obtain
\begin{align}
\Sigma(p)
&=
\frac{i g^2}{2}
\int\frac{d^4k}{(2\pi)^4}
\int d^4z_1\,e^{ik\cdot z_1}G_F(z_1)
\int d^4z_2\,e^{i(p-k)\cdot z_2}G_F(z_2)
\notag\\
&=
\frac{i g^2}{2}
\int d^4z_1 d^4z_2\,
e^{ip\cdot z_2}G_F(z_1)G_F(z_2)
\int\frac{d^4k}{(2\pi)^4}e^{ik\cdot(z_1-z_2)}
\notag\\
&=
\frac{i g^2}{2}
\int d^4z\,e^{ip\cdot z}\left(G_F(z)\right)^2 .
\label{eq:Sigma_phi3_coordinate}
\end{align}
This is the coordinate-space representation of the same bubble self-energy.

Before calculating the one-loop self-energy under light-cone smearing, to compare with our results, let us evaluate this coordinate-space integral explicitly for the usual \ac{qft} case. We consider the massless case ($m=0$) and use dimensional regularization with $d = 4 - 2\varepsilon$. To keep the coupling constant dimensionless in $d$ dimensions, we introduce an arbitrary mass scale $\mu$ and replace $g^2 \to g^2 \mu^{2\varepsilon}$. In $d$ dimensions, the massless Feynman propagator in coordinate space is given by
\begin{equation}
G_F(z) = -\frac{i}{4\pi^{d/2}}\, \Gamma\left(\frac{d}{2}-1\right) \left( \frac{1}{-z^2 + i\epsilon} \right)^{\frac{d}{2}-1}\,.
\end{equation}
Squaring this propagator yields
\begin{equation}
\left( G_F(z) \right)^2 = -\frac{1}{16\pi^d}\, \Gamma\left(\frac{d}{2}-1\right)^2 \left( \frac{1}{-z^2 + i\epsilon} \right)^{d-2}\,.
\end{equation}
Substituting this back into Eq.~\eqref{eq:Sigma_phi3_coordinate} extended to $d$ dimensions, 
we obtain
\begin{equation}
\Sigma(p) = -\frac{i g^2 \mu^{2\varepsilon}}{32\pi^d} \Gamma\left(\frac{d}{2}-1\right)^2 \int d^dz\, e^{ip\cdot z} \left( -z^2 + i\epsilon \right)^{-(d-2)}\,.   
\end{equation}

By using the standard formula,
\begin{equation}
\int d^dz \, e^{ip\cdot z} \left( -z^2 + i\epsilon \right)^{-\alpha} = -i \pi^{d/2} 2^{d-2\alpha} \frac{\Gamma(d/2-\alpha)}{\Gamma(\alpha)} \left( -p^2 - i\epsilon \right)^{\alpha-d/2}\,,   
\end{equation}
with $\alpha = d-2$, 
the one-loop self-energy is simplified to
\begin{equation}
\Sigma(p) = -\frac{g^2 \mu^{2\varepsilon}}{2 (4\pi)^{d/2}} \frac{\Gamma(d/2-1)^2 \Gamma(2-d/2)}{\Gamma(d-2)} \left( -p^2 - i\epsilon \right)^{d/2-2}\,.  
\end{equation}
Setting $d = 4 - 2\varepsilon$ and expanding around $\varepsilon = 0$, we extract the \ac{uv} divergent part and the finite result, 
\begin{equation}\label{eq:dimensional-regularization}
\Sigma(p) = -\frac{g^2}{32\pi^2} \left[ \frac{1}{\varepsilon} - \gamma_E + \ln(4\pi) - \ln \left( \frac{-p^2 - i\epsilon}{\mu^2} \right) + 2 \right] + \mathcal{O}(\varepsilon)\,,   
\end{equation}
where $\gamma_E$ is Euler's constant.
This expression precisely reproduces the standard momentum-space calculation for the massless bubble diagram, confirming that the \ac{uv} divergence arises purely from the $z^\mu \to 0$ singularity in the coordinate representation. Note that since we consider the massless scalar field, the \ac{ir} logarithmic divergence appears in the $p\to 0$ limit.

The physical mass is defined by the pole of the Dyson-resummed propagator,
\begin{equation}
D(p)=\frac{i}{p^2-m_0^2-\Sigma(p^2)+i\epsilon}\,.
\end{equation}
The pole condition is therefore
\begin{equation}
m_{\rm phys}^2-m_0^2-\Re\,\Sigma(m_{\rm phys}^2)=0\,,
\end{equation}
and, at one-loop order, this gives
\begin{equation}
m_{\rm phys}^2
=
m_0^2
+
\Re\,\Sigma(m_0^2)
+
\mathcal{O}(g^4)\,.
\end{equation}

For our present purposes, however, we are not primarily interested in the finite on-shell mass shift. Rather, our aim is to examine whether the
light-cone smearing modifies the \ac{uv} behavior of the one-loop
amplitude. We therefore denote the one-loop correction to the mass parameter by
\begin{equation}
\delta m^2
\equiv
\Re\,\Sigma(p)\,.
\label{eq:dm2_definition}
\end{equation}
This expression should be understood as the real part of the one-loop
self-energy evaluated at a given external momentum. In the strictly massless
case, we must carefully treat the limit $p^2\to 0$ because the amplitude
contains an \ac{ir} logarithm divergence. Therefore, in the following comparison, we
keep $p$ finite and focus on how the short-distance structure of the
coordinate-space integral is affected by light-cone smearing.

Now, we return to the discussion of the self-energy under light-cone smearing and
replace the products of Feynman propagator by $\langle G_F(z)^2\rangle$.  
Within the finite-resolution model, we make the replacement
$\langle\hat{\sigma}_1^2(z)\rangle\to
\Delta_{\sigma,\mathrm{eff}}^2$ and treat the effective variance as
independent of $z$ over the range relevant to the integral.
Substituting Eq.~\eqref{eq:smeared_GF_square_compact} into
Eq.~\eqref{eq:Sigma_phi3_coordinate}, we obtain
\begin{align}
\Sigma (p)
&=
\frac{i g^2}{2}
\int d^4z\,e^{ip\cdot z}
\left\langle G_F(z)^2\right\rangle
\nonumber\\
&=
\frac{i g^2}{2}
\frac{1}{64\pi^4}
\int_0^\infty ds\,
s\,
\exp\left[
-\frac{1}{2}s^2
\Delta_{\sigma,\mathrm{eff}}^2
\right]
\int d^4z\,
\exp\left[
ip\cdot z+is\sigma_0
\right].
\label{eq:Sigma_smeared_before_z_integral}
\end{align}
Using $\sigma_0=z^2/2$, the $z$-integral becomes,
\begin{equation}
I(s,p)
\equiv
\int d^4z\,
\exp\left[
ip\cdot z+\frac{i}{2}s z^2
\right].
\label{eq:Lorentzian_Gaussian_definition}
\end{equation}
With the mostly-plus convention
\begin{equation}
z^2=-(z^0)^2+|\mathbf z|^2,
\quad
p^2=-(p^0)^2+|\mathbf p|^2,
\end{equation}
this integral factorizes into
\begin{align}
I(s,p)
&=
\int dz^0\,
\exp\left[
-\frac{i}{2}s(z^0)^2-ip^0z^0
\right]
\prod_{j=1}^3
\int dz^j\,
\exp\left[
\frac{i}{2}s(z^j)^2+ip^jz^j
\right].
\label{eq:Lorentzian_Gaussian_factorized}
\end{align}
For $s>0$, with the usual Feynman prescription, we obtain
\begin{align}
\int dz^0\,
\exp\left[
-\frac{i}{2}s(z^0)^2-ip^0z^0
\right]
&=
\left(\frac{2\pi}{s}\right)^{1/2}
e^{-i\pi/4}
\exp\left[
\frac{i(p^0)^2}{2s}
\right],
\\
\int dz^j\,\exp\left[
\frac{i}{2}s(z^j)^2+ip^jz^j
\right]
&=\left(\frac{2\pi}{s}\right)^{1/2}
e^{i\pi/4}
\exp\left[
-\frac{i(p^j)^2}{2s}
\right]\,.
\end{align}
By using the above expressions, we find
\begin{align}
I(s,p)
&=i\frac{(2\pi)^2}{s^2}
\exp\left[-\frac{i}{2s}\left(p^2-i0\right)
\right].
\label{eq:Lorentzian_Gaussian_result}
\end{align}
The infinitesimal $i0$ specifies the convergence of the oscillatory integral near $s=0$.

Using Eq.~\eqref{eq:Lorentzian_Gaussian_result} in
Eq.~\eqref{eq:Sigma_smeared_before_z_integral}, we obtain
\begin{align}
\Sigma (p)
&=
\frac{i g^2}{2}
\frac{1}{64\pi^4}
\int_0^\infty ds\,
s\,
\exp\left[
-\frac{1}{2}s^2
\Delta_{\sigma,\mathrm{eff}}^2
\right]
\left[
i\,\frac{(2\pi)^2}{s^2}
\exp\left[
-\frac{i}{2s}
\left(p^2-i0\right)
\right]
\right]
\nonumber\\
&=
-\frac{g^2}{32\pi^2}
\int_0^\infty
\frac{ds}{s}\,
\exp\left[
-\frac{1}{2}s^2
\Delta_{\sigma,\mathrm{eff}}^2
-\frac{i}{2s}
\left(p^2-i0\right)
\right].
\label{eq:Sigma_smeared_final_complex}
\end{align}
Therefore the real part is
\begin{equation}
\Re\,\Sigma (p)
=
-\frac{g^2}{32\pi^2}
\int_0^\infty
\frac{ds}{s}\,
\exp\left[
-\frac{1}{2}s^2
\Delta_{\sigma,\mathrm{eff}}^2
\right]
\cos\left(
\frac{p^2}{2s}
\right)
\label{eq:Re_Sigma_smeared_final}\,, 
\end{equation}
where the large-$s$ region of Eq.~\eqref{eq:Re_Sigma_smeared_final} is
exponentially suppressed by $\exp\left[
-\frac{1}{2}s^2
\Delta_{\sigma,\mathrm{eff}}^2
\right]$.
At fixed $\Delta_{\sigma,\mathrm{eff}}^2>0$, this exponential suppression is the direct imprint of the coarse-grained smearing of the light-cone singularity and makes the ultraviolet-sensitive short-distance part finite. It is also important to examine $s\to 0$. For $p^2\neq 0$, the factor $\cos \left(p^2/2s\right)$ oscillates infinitely rapidly near $s=0$. Although the integral is not absolutely convergent there,
it is well defined as an oscillatory integral, or the original $i0$ prescription. By using $u=|p^2|/(2s)$, the small-$s$ part is reduced to
the integral
\begin{equation}
\int_0
\frac{ds}{s}\,
\cos\left(
\frac{p^2}{2s}
\right) \to 
\int^\infty \frac{du}{u}\cos u \,,   
\end{equation}
which is conditionally convergent. 
Thus, at fixed nonzero effective variance, the off-shell amplitude with $p^2\neq 0$ is finite.

However, for $p^2=0$ this oscillatory factor becomes unity and we see the 
logarithmic divergence at $s\to0$,
\begin{equation}
\Re\,\Sigma (0)
=
-\frac{g^2}{32\pi^2}
\int_0^\infty
\frac{ds}{s}\,
\exp\left[
-\frac{1}{2}s^2
\Delta_{\sigma,\mathrm{eff}}^2
\right].
\label{eq:Re_Sigma_smeared_p_zero}
\end{equation}
Introducing a lower cutoff $s_{\rm min}$, we find
\begin{align}
\Re\,\Sigma (0)
&=-\frac{g^2}{32\pi^2}
\int_{s_{\rm min}}^\infty
\frac{ds}{s}\,
\exp\left[
-\frac{1}{2}s^2
\Delta_{\sigma,\mathrm{eff}}^2
\right]
\nonumber\\
&=
-\frac{g^2}{64\pi^2}
E_1\left(
\frac{1}{2}
\Delta_{\sigma,\mathrm{eff}}^2
s_{\rm min}^2
\right),
\label{eq:Re_Sigma_smeared_cutoff_E1}
\end{align}
where
\begin{equation}
E_1(x)\equiv \int_x^\infty dt\,\frac{e^{-t}}{t}
\end{equation}
is the exponential integral.  For small $s_{\rm min}$, we get
\begin{align}
\Re\,\Sigma (0)
&\simeq
-\frac{g^2}{64\pi^2}
\left[
-\gamma_E
-\ln\left(
\frac{1}{2}
\Delta_{\sigma,\mathrm{eff}}^2
s_{\rm min}^2
\right)
\right]
+\mathcal{O}(s_{\rm min}^2)\,.
\label{eq:Re_Sigma_smeared_cutoff_log}
\end{align}
This logarithmic divergence should not be interpreted as the original local
\ac{uv} divergence associated with $z^\mu\to0$. At fixed nonzero effective variance, the ultraviolet-sensitive short-distance part is finite, as shown by
Eq.~\eqref{eq:Re_Sigma_smeared_final}. Instead, the divergence in
Eq.~\eqref{eq:Re_Sigma_smeared_p_zero} arises from the $s\to0$ endpoint of the
auxiliary representation, or the \ac{ir} sector of the  massless propagator.

To explicitly extract the asymptotic behavior and clarify the role of the effective smearing parameter $\Delta_{\sigma,\mathrm{eff}}^2$, it is instructive to examine $\Sigma(p)$ for $p^2 \neq 0$ in the limit $\Delta_{\sigma,\mathrm{eff}}^2 \to 0$. The complete integral, with the original $i0$ prescription restored and without separating the real part, is
\begin{equation}
\Sigma(p) = -\frac{g^2}{32\pi^2} \int_0^\infty \frac{ds}{s} \exp\left[ -\frac{1}{2}s^2 \Delta_{\sigma,\mathrm{eff}}^2 -\frac{i}{2s} (p^2-i0) \right].
\end{equation}
To evaluate this, we define the parameters $A = \frac{1}{2}\Delta_{\sigma,\mathrm{eff}}^2$ and $B = \frac{i}{2}(p^2-i0)$ to simplify the notation, reducing the integral to the form
\begin{equation}
I(A,B) = \int_0^\infty \frac{ds}{s} \exp\left(-A s^2 - \frac{B}{s}\right).
\end{equation}
By utilizing the Mellin-Barnes representation for the exponential function, $\exp(-B/s) = \frac{1}{2\pi i} \int_{c-i\infty}^{c+i\infty} dz \, \Gamma(z) (B/s)^{-z}$ with $c>0$, we can perform the $s$-integration first,
\begin{equation}
I(A,B) = \frac{1}{2\pi i} \int_{c-i\infty}^{c+i\infty} dz \, \Gamma(z) B^{-z} \int_0^\infty ds \, s^{z-1} e^{-A s^2}.
\end{equation}
The integral over $s$ yields $\frac{1}{2} A^{-z/2} \Gamma(z/2)$. Combining these, we obtain a contour integral representation,
\begin{equation}
I(A,B) = \frac{1}{4\pi i} \int_{c-i\infty}^{c+i\infty} dz \, \Gamma(z) \Gamma\left(\frac{z}{2}\right) (B\sqrt{A})^{-z}.
\end{equation}
The asymptotic expansion for small $A$ is determined by the poles of the integrand in the left half of the complex $z$-plane. The dominant contribution arises from the double pole at $z=0$. Expanding the integrand around $z=0$ using $\Gamma(z) \simeq 1/z - \gamma_E$ and $(B\sqrt{A})^{-z} \simeq 1 - z \ln(B\sqrt{A})$, we can extract the residue. Multiplying by the overall prefactor $1/2$ originating from the contour integration, the asymptotic form becomes
\begin{align}
I(A,B) &\simeq \frac{1}{2} \text{Res}_{z=0} \left[ \Gamma(z) \Gamma\left(\frac{z}{2}\right) (B\sqrt{A})^{-z} \right] \nonumber \\
&= -\frac{1}{2}\ln A - \ln B - \frac{3}{2}\gamma_E \,.
\end{align}
Substituting back the physical parameters $A = \frac{1}{2}\Delta_{\sigma,\mathrm{eff}}^2$ and $B = \frac{i}{2}(p^2-i0)$, we have
\begin{equation}
I \simeq -\frac{1}{2}\ln\left(\frac{1}{2}\Delta_{\sigma,\mathrm{eff}}^2\right) - \ln\left[\frac{i}{2}(p^2-i0)\right] - \frac{3}{2}\gamma_E \,.
\end{equation}
To compare the dimensional regularization results, we introduce an arbitrary mass scale $\mu$ to construct dimensionless arguments for the logarithms. Using the principal value $\ln(i) = i\pi/2$ and the analytic continuation $\ln(p^2-i0) = \ln(-p^2+i0) - i\pi$, we can find the asymptotic result,
\begin{equation}\label{eq:Sigma_smeared_final}
\Sigma(p) \simeq -\frac{g^2}{32\pi^2} \left[ -\frac{1}{2}\ln(\mu^4\Delta_{\sigma,\mathrm{eff}}^2) - \frac{3}{2}\gamma_E + \frac{3}{2}\ln 2 - \ln\left(\frac{-p^2+i0}{\mu^2}\right) + i\frac{\pi}{2} \right].
\end{equation}
This expression takes a form similar to the standard result~\eqref{eq:dimensional-regularization} obtained via dimensional regularization. For a prescribed nonzero $\Delta_{\sigma,\mathrm{eff}}^2$, the logarithm shows that the smearing width plays a regulator role: the usual ultraviolet singularity is recovered as $\Delta_{\sigma,\mathrm{eff}}^2\to0$, while the remaining finite terms correctly reproduce the expected logarithmic structures, $-\ln((-p^2+i0)/\mu^2)$, of the standard
one-loop self-energy. This demonstrates that the smearing procedure consistently regularizes
the short-distance singularities while perfectly preserving the exact physical long-distance behavior and analytic structure of the theory is retained.

The behavior of the point-split variance can be estimated from the first-order correction to Synge's world function,
\begin{equation}
\sigma_1 \sim \frac{1}{2}h_{\mu\nu}z^\mu z^\nu \,.
\end{equation}
Strictly speaking, the graviton field is an operator and its
short-distance two-point function takes the form
\begin{equation}
\langle
\hat h_{\mu\nu}(x)\hat h_{\rho\sigma}(x')
\rangle
\sim
\frac{1}{M_{\rm pl}^2}
\frac{\Pi_{\mu\nu\rho\sigma}}{z^2}\,.
\end{equation}
where $\Pi_{\mu\nu\rho\sigma}$ denotes a dimensionless tensor structure
depending on the gauge choice and the polarization sum. Substituting this
short-distance behavior into the estimate of $\sigma_1$, we obtain
\begin{equation}
\langle \hat{\sigma}_1^2(x,x')\rangle
\sim
z^\mu z^\nu z^\rho z^\sigma
\langle
\hat h_{\mu\nu}(x)\hat h_{\rho\sigma}(x')
\rangle
\sim
\frac{\ell^2}{M_{\rm pl}^2},
\label{eq:point-split-variance-estimate}
\end{equation}
where $\ell$ is a characteristic length scale. 
In the strict point-splitting limit, $\ell$ would be identified with the separation between
$x$ and $x'$, and hence $\ell\to 0$ as $x'\to x$.  In that case the leading-order
variance $\langle \hat{\sigma}_1^2(x,x')\rangle$ formally vanishes in the coincidence limit.

Equation~\eqref{eq:point-split-variance-estimate} also exposes a limitation of the present perturbative treatment. If $\ell$ is identified with the point-splitting separation, then $\ell\to0$ as $x'\to x$, and the leading-order variance $\langle\hat{\sigma}_1^2(x,x')\rangle$ vanishes. Consequently, within quantum field theory formulated on a smooth continuum background, 
at the first-order correction of the metric perturbations, the present formalism does not remove the coincident-point ultraviolet singularity. 
\footnote{
However, when higher-order corrections of the spacetime perturbation beyond first order are taken into account in Synge's world function, it is not clear whether the coincident singularity can be removed. This is because $\langle \hat{\sigma}_2 \rangle$ can in fact be interpreted as finite.}
This is consistent with the original analysis of Ref.~\cite{Ford:1994cr}, in which the one-loop electron self-energy retains coincident-point singularities.

We may nevertheless investigate the consequences of finite spacetime resolution by introducing a nonzero coarse-graining scale $\ell$ as an additional phenomenological input. Under this prescription, we estimate
\begin{equation}
\Delta_{\sigma,\mathrm{eff}}
\equiv
\sqrt{\Delta_{\sigma,\mathrm{eff}}^2}
\sim
\frac{\ell}{M_{\rm pl}}
=
\ell\,l_{\rm pl}\,,
\label{eq:effective-smearing-width}
\end{equation}
where $l_{\rm pl}$ is the Planck length. Since Synge's world function has dimensions of length squared, $\Delta_{\sigma,\mathrm{eff}}$ is a width in the world function rather than a length. At fixed $\Delta_{\sigma,\mathrm{eff}}^2>0$, the ultraviolet-sensitive short-distance part of the off-shell bubble and the tadpole expressions is finite within the effective model. This conditional finiteness should not, however, be interpreted as demonstrating that metric fluctuations within the continuum formulation dynamically eliminate ultraviolet divergences.

The microscopic origin of such a finite resolution scale, and whether it requires a modification or replacement of the smooth-manifold description of spacetime, lie beyond the scope of the present analysis. In the numerical estimates below, the choice $\ell=l_{\rm pl}$ is made only for illustration and is not determined by the present formalism.

In addition to this vacuum contribution, primordial gravitons generated
during inflation give a contribution dominated by
long-wavelength modes. Restricting the modes to the superhorizon
range within a finite interval of comoving momenta, we estimate
\begin{equation}
\langle \hat h^2\rangle_{\rm inf}
\sim
\frac{2H^2}{\pi^2M_{\rm pl}^2}N_{\rm eff}\,.
\end{equation}
where $H$ is the Hubble rate during inflation and $N_{\rm eff}$ denotes the effective number of e-folds. See Appendix~\ref{app:squeezed-gravitons-cosmology} for details.
For a coarse-graining scale $\ell$, this gives the long-wavelength contribution
\begin{equation}
\Delta_{\sigma,\mathrm{inf}}^2
\equiv
\langle \hat{\sigma}_1^2\rangle_{\rm inf}
\sim
\ell^4\langle \hat h^2\rangle_{\rm inf}
\sim
\ell^4
\frac{2H^2}{\pi^2M_{\rm pl}^2}N_{\rm eff}.
\end{equation}
Under the same phenomenological prescription, the effective short-distance width is
estimated as
\begin{equation}
\Delta_{\sigma,\mathrm{eff}}^2
\sim
\frac{\ell^2}{M_{\rm pl}^2}
=\ell^2 l_{\rm pl}^2\,.
\end{equation}
Hence their ratio is
\begin{equation}\label{eq:ratio-inf-vacuum}
\frac{
\Delta_{\sigma,\mathrm{inf}}^2}{
\Delta_{\sigma,\mathrm{eff}}^2
}\sim
\frac{2}{\pi^2}H^2\ell^2 N_{\rm eff}\,.
\end{equation}

Using the logarithmic dependence of the smeared self-energy on
$\Delta_{\sigma,\mathrm{eff}}^2$, the change induced by the inflationary
gravitons can be estimated as
\begin{equation}
\Sigma_{\rm inf}(p)-\Sigma(p)
\simeq
\frac{g^2}{64\pi^2}
\ln\left[\frac{\Delta_{\sigma,\mathrm{eff}}^2
+\Delta_{\sigma,\mathrm{inf}}^2
}{\Delta_{\sigma,\mathrm{eff}}^2}
\right]\,,
\label{eq:Sigma_inf_minus_Mink}
\end{equation}
where $\Sigma (p)$ denotes the smeared self-energy evaluated with the
prescribed baseline width $\Delta_{\sigma,\mathrm{eff}}^2$, while
$\Sigma_{\rm inf}(p)$ includes the additional contribution
from primordial gravitons.
Equivalently, we obtain
\begin{equation}
\Sigma_{\rm inf}(p)-\Sigma (p)
\sim
\frac{g^2}{64\pi^2}\ln\left[
1+ \frac{2}{\pi^2}H^2\ell^2 N_{\rm eff}
\right]\,.
\end{equation}
For $\Delta_{\sigma,\mathrm{inf}}^2
\ll \Delta_{\sigma,\mathrm{eff}}^2$, 
this reduces to
\begin{equation}
\Sigma_{\rm inf}(p)-\Sigma(p)
\sim
\frac{g^2}{64\pi^2}
\frac{2}{\pi^2}H^2\ell^2 N_{\rm eff}\,.
\end{equation}
Thus, the primordial-graviton contribution gives a finite and \ac{ir} correction to the light-cone smearing width, and
hence to the one-loop self-energy. This correction should be
distinguished from the ordinary \ac{qft}. 
The latter is controlled by the short-distance vacuum
fluctuation, whereas the inflationary contribution originates from primordial gravitons.

For illustration, if we take the characteristic length to be the Planck length, $\ell=l_{\rm pl}=M_{\rm pl}^{-1}$, the correction becomes
\begin{equation}
\Sigma_{\rm inf}(p)-\Sigma (p)
\simeq
\frac{g^2}{64\pi^2}
\frac{2}{\pi^2}
\left(\frac{H}{M_{\rm pl}}\right)^2
N_{\rm eff}\,.
\end{equation}
For $H=10^{13}\,{\rm GeV}$, $N_{\rm eff}\sim 60$, and the reduced Planck mass
$M_{\rm pl}\simeq 2.4\times10^{18}\,{\rm GeV}$, we find
\begin{equation}
\Sigma_{\rm inf}(p)-\Sigma(p)
\sim
\frac{g^2}{64\pi^2}
\frac{2}{\pi^2}
\left(\frac{H}{M_{\rm pl}}\right)^2
N_{\rm eff}
\sim
3.3\times10^{-13}g^2.
\end{equation}
The primordial-graviton contribution produces a tiny but finite one-loop correction to the mass, of $\mathcal{O}(10^{-13}g^2)$. This correction, however, should not be regarded as a directly observable mass shift by itself. In the approximation used here, it is essentially independent of momentum and can therefore be absorbed into the renormalized mass parameter. In this respect, it differs from the usual running of couplings, where the scale dependence of scattering amplitudes can lead to observable consequences. To extract a physical signature, we should identify an observable for which the primordial-graviton contribution is not  eliminated by the parameter redefinition. It is also important to avoid interpreting the effect as a direct difference between measurements performed in regions with primordial gravitational waves and those performed in regions without them. Such a comparison is not available in local experiments. For primordial gravitational waves generated during inflation, the long-wavelength modes are effectively quasi-static over laboratory scales, while their observable influence must be treated statistically on cosmological scales. Within this framework, the correction discussed above may still leave a testable imprint.

Although the estimated correction is extremely small, its relative size compared with an ordinary one-loop coefficient is of order $10^{-10}$. Since a one-loop QED contribution to a lepton anomalous magnetic moment is of order $10^{-3}$, the corresponding absolute change would be of order $10^{-13}$. Such a shift is far below the present sensitivity of muon $g{-}2$ measurements, but it is closer to the level probed in precision measurements of the electron magnetic moment. Precision $g{-}2$ experiments may therefore offer a possible route to testing light-cone smearing effects induced by primordial gravitational waves, although a detailed analysis would be required.

\subsection{$\phi^4$ theory: tadpole self-energy and coincident singularity}
\label{subsubsec:phi4_tadpole}

Next, we consider the quartic interaction
\begin{equation}
\mathcal{L}_{\rm int}
=
-\frac{\lambda}{4!}\phi^4 .
\end{equation}
At one loop, the self-energy is given by the tadpole diagram.  With the
same convention for the propagator~\eqref{eq:GF_fourier}, the one-loop 
contribution is
\begin{equation}
\Sigma_{\rm tad}
=
\frac{i\lambda}{2}
\int\frac{d^4k}{(2\pi)^4}
\frac{1}{k^2-m^2+i\epsilon}
=
\frac{i\lambda}{2}\,G_F(0)\,,
\label{eq:Sigma_phi4_tadpole}
\end{equation}
where the factor $1/2$ is the symmetry factor of the tadpole diagram.  In
contrast to the $\phi^3$ bubble diagram, there is only one internal propagator.
Therefore the one-loop correction is momentum independent and directly probes
the coincident-point singularity of the Feynman propagator.

To make the \ac{uv} behavior explicit, let us first recall the usual
cutoff result after Wick rotation.  The corresponding correction to the mass
parameter is
\begin{align}
\delta m^2
&= \frac{\lambda}{2}
\int^{\Lambda_{\rm UV}}
\frac{d^4k_E}{(2\pi)^4}
\frac{1}{k_E^2+m^2}
\nonumber\\
&= \frac{\lambda}{32\pi^2}
\left[\Lambda_{\rm UV}^2 -
m^2\ln\left(1+\frac{\Lambda_{\rm UV}^2}{m^2}\right)
\right].
\label{eq:phi4_tadpole_cutoff}
\end{align}
Thus the tadpole diagram contains a quadratic \ac{uv} divergence.
In the massless limit this becomes
\begin{equation}
\delta m^2
\simeq
\frac{\lambda}{32\pi^2}\Lambda_{\rm UV}^2 .
\label{eq:phi4_tadpole_massless_cutoff}
\end{equation}
In dimensional regularization, the strictly massless tadpole integral is set to zero. However, this should not be interpreted as the absence of \ac{uv} sensitivity, because dimensional regularization removes power divergences. For the present purpose, where we aim to identify the short-distance singularity, the cutoff expression in Eq.~\eqref{eq:phi4_tadpole_cutoff} is more transparent.

We now apply the light-cone smearing to the tadpole contribution. Using the smeared Feynman propagator at the coincidence point, we obtain
\begin{align}
\left\langle G_F(0)\right\rangle
&= -\frac{1}{8\pi^2}
\int_0^\infty du\,
\exp\left[
-\frac{1}{2}u^2
\Delta_{\sigma,\mathrm{eff}}^2
\right]
\nonumber\\
&=
-\frac{1}{8\pi^2}
\sqrt{
\frac{\pi}{2\Delta_{\sigma,\mathrm{eff}}^2}
}.
\label{eq:smeared_GF_coincident}
\end{align}
At fixed nonzero $\Delta_{\sigma,\mathrm{eff}}^2$, the coincident-point expression in this coarse-grained model is finite. The
smeared tadpole contribution becomes
\begin{equation}
\Sigma_{\rm tad}
=\frac{i\lambda}{2}
\left\langle G_F(0)\right\rangle
=
-\frac{i\lambda}{16\pi^2}
\sqrt{
\frac{\pi}{2\Delta_{\sigma,\mathrm{eff}}^2}
}.
\label{eq:Sigma_tadpole_smeared}
\end{equation}
Equivalently, the one-loop mass correction is of order
\begin{equation}
\delta m^2_{\rm tad}
\sim
\frac{\lambda}{16\pi^2}
\sqrt{
\frac{\pi}{2\Delta_{\sigma,\mathrm{eff}}^2}
}.
\label{eq:dm2_tadpole_smeared}
\end{equation}
This expression has the correct mass dimension, since
$\Delta_{\sigma,\mathrm{eff}}^2$ has dimension of length to the fourth power. It is useful to compare this result with Eq.~\eqref{eq:phi4_tadpole_massless_cutoff}. At fixed nonzero effective width, the coarse-grained expression is formally analogous to replacing the quadratic \ac{uv} cutoff by
\begin{equation}
\Lambda_{\rm UV}^2
\sim
\frac{1}{\sqrt{\Delta_{\sigma,\mathrm{eff}}^2}}\,.
\label{eq:effective-cutoff}
\end{equation}
This analogy characterizes the coarse-grained model; it should not be interpreted as an intrinsic removal of the ultraviolet divergence by the leading-order continuum theory.

We can also estimate how the tadpole contribution is modified by
primordial gravitons generated during inflation.  As discussed above, the smearing width receives both the short-distance vacuum contribution and the long-wavelength inflationary contribution.  
Using the smeared tadpole result in Eq.~\eqref{eq:dm2_tadpole_smeared}, the tadpole mass correction in the presence of primordial gravitons becomes
\begin{equation}
\delta m^2_{\rm tad,inf}
\sim
\frac{\lambda}{16\pi^2}
\sqrt{\frac{\pi}{2\left(
\Delta_{\sigma,\mathrm{eff}}^2
+\Delta_{\sigma,\mathrm{inf}}^2
\right)}
}\,.
\label{eq:dm2_tadpole_inf}
\end{equation}
Hence, the change relative to the vacuum contribution is
\begin{align}
\delta m^2_{\rm tad,inf}
- \delta m^2_{\rm tad}
&\sim \frac{\lambda}{16\pi^2}
\sqrt{\frac{\pi}{2}}
\left[\frac{1}{\sqrt{
\Delta_{\sigma,\mathrm{eff}}^2
+\Delta_{\sigma,\mathrm{inf}}^2}}
-\frac{1}{\sqrt{\Delta_{\sigma,\mathrm{eff}}^2}}
\right]
\nonumber\\
&=
\delta m^2_{\rm tad}\left[\left(
1+\frac{\Delta_{\sigma,\mathrm{inf}}^2
}{\Delta_{\sigma,\mathrm{eff}}^2}\right)^{-1/2}-1\right].
\label{eq:dm2_tadpole_inf_minus_vac}
\end{align}
For
$\Delta_{\sigma,\mathrm{inf}}^2
\ll
\Delta_{\sigma,\mathrm{eff}}^2$,
this reduces to
\begin{equation}
\delta m^2_{\rm tad,inf}
-
\delta m^2_{\rm tad}
\sim
-\frac{1}{2}\,
\delta m^2_{\rm tad}
\frac{
\Delta_{\sigma,\mathrm{inf}}^2
}{
\Delta_{\sigma,\mathrm{eff}}^2
}\, .
\label{eq:dm2_tadpole_inf_small}
\end{equation}
Using Eq.~\eqref{eq:ratio-inf-vacuum},
we obtain
\begin{equation}
\frac{
\delta m^2_{\rm tad,inf}
-\delta m^2_{\rm tad}}{\delta m^2_{\rm tad}}
\sim
-\frac{1}{\pi^2}\left(\frac{H}{M_{\rm pl}}\right)^2
N_{\rm eff}\,.
\label{eq:relative_tadpole_inf}
\end{equation}
where we take $\ell=l_{\rm pl}=M_{\rm pl}^{-1}$.
Thus, in the $\phi^4$ tadpole diagram, the primordial-graviton contribution reduces the magnitude of the tadpole correction.  
For $H=10^{13}\,{\rm GeV}$, $N_{\rm eff}\sim60$, and
$M_{\rm pl}\simeq2.4\times10^{18}\,{\rm GeV}$, this gives
\begin{equation}
\frac{
\delta m^2_{\rm tad,inf}
-\delta m^2_{\rm tad}}{\delta m^2_{\rm tad}}
\sim -1.1\times10^{-10}.
\label{eq:relative_tadpole_inf_numerical}
\end{equation}
Therefore the inflationary graviton contribution induces a tiny relative shift of order $10^{-10}$ in the smeared tadpole correction.  However, as in the bubble diagram, this should not be immediately interpreted as a directly observable mass shift.  The tadpole contribution is momentum independent and
can be absorbed into the renormalized mass parameter. To extract a physical signature, we must identify an observable in
which the primordial-graviton contribution cannot be removed by a
redefinition of the renormalized mass.  In particular, it should not be interpreted as a direct difference between local measurements performed in regions with and without primordial gravitational waves. For inflationary gravitons, the relevant long-wavelength modes are effectively quasi-static on laboratory scales, while their observable effects should be treated statistically on cosmological scales.  Within this framework, the small shift
in the smearing width may nevertheless provide a possible imprint of
primordial graviton fluctuations.

\section{Conclusion and Discussion}
\label{sec:conclusion}

In this work, we have investigated how quantum fluctuations of the
gravitational field smear the classical light-cone structure and how this
smearing modifies the short-distance behavior of perturbative \ac{qft}.  Starting from the operator-valued perturbation of Synge's world function,
$\hat{\sigma}_1$, induced by graviton fluctuations $\hat h_{\mu\nu}$, we evaluated the expectation value of the retarded Green's function for several graviton states. The classical light-cone singularity $\delta(\sigma_0)$ is replaced by a Gaussian distribution whose width is controlled by the variance $\langle \hat{\sigma}_1^2\rangle$.

A central result of our analysis is that the smearing effect depends on the quantum state of the gravitational field. Coherent states, which correspond to classical gravitational waves, mainly shift the location of the light cone. By contrast, squeezed vacuum states change the variance itself. By squeezing and anti-squeezing, we mean, respectively, phase choices for which the mode contribution $|g_k|^2$ is smaller or larger than the vacuum contribution $|f_k|^2$. Thus squeezing suppresses the mode contribution to the light-cone smearing, whereas anti-squeezing enhances it. This distinction is important because primordial gravitons generated during inflation are naturally described by highly squeezed states and can therefore leave an imprint different from that of a classical stochastic
background.

It is important, however, to clarify the interpretation of the smearing width. At leading order in the metric perturbation, $\hat{\sigma}_1$ is proportional to the separation between the two spacetime points. Therefore, in the strict point-splitting limit $x'\to x$, the leading-order variance $\langle\hat{\sigma}_1^2(x,x')\rangle$ formally vanishes. This is consistent with the original analysis of Ref.~\cite{Ford:1994cr}, where the light-cone singularity is smeared for distinct points, but the one-loop electron self-energy still retains residual coincident-point singularities. At the leading-order metric perturbation, this formalism therefore does not derive a nonzero coincidence width, a fundamental minimal length, or an ultraviolet completion. For the loop calculations, we instead introduce $\Delta_{\sigma,\mathrm{eff}}^2>0$ as a prescribed phenomenological coarse-grained variance.

At fixed $\Delta_{\sigma,\mathrm{eff}}^2$, the ultraviolet-sensitive part of the off-shell $\phi^3$ bubble is finite, while the asymptotic expansion reproduces the expected logarithmic momentum dependence. For the $\phi^4$ tadpole, the coarse-grained coincident-point expression is finite and is formally analogous to a quadratic cutoff scale $\Lambda_{\rm UV}^2\sim1/\sqrt{\Delta_{\sigma,\mathrm{eff}}^2}$. These results show that a prescribed nonzero smearing width can act as an effective ultraviolet regulator for the singular short-distance part of perturbative \ac{qft}. This statement is conditional: if the smearing width is taken to zero, one recovers the formal local-\ac{qft} limit and the usual coincident-point singularities reappear.

Furthermore, by including the contribution of long-wavelength primordial gravitons produced during inflation, we estimated the relative correction to the smeared self-energy to be of order $(H/M_{\rm pl})^2 N_{\rm eff}$. Conditional on the illustrative choice $\ell=l_{\rm pl}$, the parameters $H\sim 10^{13}\,{\rm GeV}$ and $N_{\rm eff}\sim 60$ give a relative shift of order $10^{-10}$ in the tadpole estimate. Although this magnitude is well below the present sensitivity of most precision experiments, it is not entirely beyond reach; the implied modification of one-loop QED effects would translate into a shift in the lepton anomalous magnetic moment that approaches the precision level achieved in electron $g{-}2$ measurements. We emphasize that this correction is essentially momentum independent in the present approximation and thus largely absorbed into the renormalized mass parameter.
Thus, whether or not the light-cone smearing effect can be observed remains a critical question.

Despite this limitation, the framework presented here demonstrates how graviton fluctuations modify light-cone propagation, how classical and non-classical graviton states leave qualitatively distinct imprints, and how a finite-resolution effective propagator changes the short-distance structure of loop amplitudes.

\section*{Acknowledgments}
H. M. would like to thank Shinji Mukohyama and Sugumi Kanno for helpful discussions on the quantum nature of primordial gravitational waves, and Tomohiro Fujita for valuable discussions on light-cone smearing. This work was supported by JSPS KAKENHI Grant No. JP23K13100.

\appendix

\section{Squeezed gravitons in inflationary cosmology}
\label{app:squeezed-gravitons-cosmology}

In this appendix, we review the general properties of primordial gravitons in an expanding \ac{flrw} universe and show that cosmological evolution naturally leads to a squeezed state of these gravitons~\cite{Maggiore:1999vm,Kanno:2018cuk,Kanno:2019gqw}.

We consider a spatially flat \ac{flrw} background with tensor perturbations,
\begin{equation}
\mathrm{d}s^2
=
-\mathrm{d}t^2
+
a^2(t)\bigl(\delta_{ij}+h_{ij}\bigr)\mathrm{d}x^i\mathrm{d}x^j,
\end{equation}
where $a(t)$ is the scale factor and $h_{ij}$ denotes the transverse and traceless tensor perturbation satisfying
\begin{equation}
\partial_i h_{ij}=0,
\quad
h_{ii}=0,
\quad
|h_{ij}|\ll 1.
\end{equation}

At quadratic order, the action for the tensor perturbation is
\begin{equation}
S^{(2)}=
\frac{M_{\rm pl}^2}{8} \int \mathrm{d}t \mathrm{d}^3x 
a^3\biggl[\dot{h}^{i j} \dot{h}_{i j} - h^{i j}\Delta h_{i j}\biggr]\,,
\end{equation}
where $M_{\rm pl}^2=1/(8\pi G)$ and $\Delta=g^{i j}\nabla_i\nabla_j=a(t)^{-2}\gamma^{i j}\nabla_i\nabla_j$ is the Laplacian associated with the spatial metric $g_{i j}$. 
Introducing the conformal time $\eta$ by
$\mathrm{d}\eta=\mathrm{d}t/a$, the tensor perturbation can be expanded as
\begin{equation}
h_{ij}(\eta,\bm x)
=\frac{\sqrt{2}}{M_{\rm pl}}
\sum_{s=+,\times}
\int\frac{\mathrm{d}^3k}{(2\pi)^3}\,
h_{\bm k}^s(\eta)\,
e^{i\bm k\cdot\bm x}\,
p_{ij}^s(\bm k),
\label{eq:fourier-tensor}
\end{equation}
where $p_{ij}^s(\bm k)$ is the polarization tensor satisfying
\begin{equation}
p_{ii}^s(\bm k)=0,
\quad
k_i p_{ij}^s(\bm k)=0,
\quad
p_{ij}^{*s}(\bm k)\,p_{ij}^{s'}(\bm k)=2\delta^{ss'}.
\end{equation}
Due to the reality condition of $h_{ij}$, 
the Fourier modes satisfy $h^{s*}_{\bm{k}}(\eta)=h^s_{-\bm{k}}(\eta)$.
Defining the canonically normalized field as $\tilde{h}^s_{\bm{k}}(\eta) \equiv a(\eta) h^s_{\bm{k}}(\eta)$,
it obeys the following equation of motion:
\begin{equation}
\tilde{h}_{\bm k}^{\prime\prime s}+\left(
k^2-\frac{a''}{a}\right) \tilde{h}^s_{\bm k}=0\,,
\label{eq:eom}
\end{equation}
where the prime denotes the derivative with respect to the conformal time $\eta$.

We now quantize the tensor perturbation. 
Promoting the tensor field $\tilde{h}^s_{\bm{k}}(\eta)$ to a quantum operator $\hat{\tilde{h}}^s_{\bm{k}}(\eta)$, we can expand it in terms of the initial mode functions $f_k(\eta)$ as
\begin{equation}
\hat{\tilde{h}}^s_{\bm k}(\eta) = f_k(\eta)\hat{a}^s_{\bm k} + f_k^{*}(\eta)\hat{a}_{-\bm k}^{s \dagger}\,,
\end{equation}
where the creation and annihilation operators satisfy the standard commutation relations $[\hat{a}^s_{\bm k} , \hat{a}_{\bm p}^{s'\dagger}] = (2\pi)^3 \delta^{ss'} \delta^{(3)}({\bm k}-{\bm p})$.
Alternatively, the same operator can be expanded in terms of the late-time mode functions $g_{k}(\eta)$ as
\begin{equation}
\hat{\tilde{h}}^s_{\bm k}(\eta) = g_k(\eta)\hat{b}^s_{\bm k} + g_k^{*}(\eta)\hat{b}_{-\bm k}^{s \dagger}\,,
\end{equation}
where the creation and annihilation operators satisfy $[\hat{b}^s_{\bm k} , \hat{b}_{\bm p}^{s'\dagger}]= (2\pi)^3 \delta^{ss'} \delta^{(3)}({\bm k}-{\bm p})$.

The initial vacuum state $|0\rangle_{a}$ (e.g., the Bunch-Davies vacuum) and the late-time vacuum state $|0\rangle_{b}$ 
are defined respectively by $\hat{a}^s_{\bm k}|0\rangle_{a}=0$ and $\hat{b}^s_{\bm k}|0\rangle_{b}=0$. 
The relation between the two operators is governed by the Bogoliubov transformation:
\begin{align}\label{eq:Bogoliubov-transformation}
\hat{b}^s_{\bm k}&=
\alpha_k^* \hat{a}^s_{\bm k} - \beta_k^* \hat{a}_{-\bm k}^{s\dagger}\,, \\
\hat{b}_{-\bm k}^{s\dagger}&=-\beta_k \hat{a}^s_{\bm k} + \alpha_k \hat{a}_{-\bm k}^{s\dagger} \,.
\end{align}
To be consistent with the definition of the two-mode squeezing operator in the previous section, we parameterize the Bogoliubov coefficients as
\begin{equation}
\alpha_k = \cosh r_k\,, \quad
\beta_k = e^{-i\theta_k}\sinh r_k\,.
\end{equation}
The corresponding graviton
occupation number is
\begin{equation}
n_k=|\beta_k|^2=\sinh^2 r_k.
\end{equation}
From the viewpoint of the late-time observer associated with the
$\hat{b}_{\bm k}^s$ operators, the initial vacuum $\ket{0}_a$ is a two-mode squeezed state. More explicitly, we have
\begin{equation}
|0\rangle_{a}
= \prod_{s,\bm k} \frac{1}{\cosh r_k} \sum_{n=0}^\infty (-1)^ne^{-in\theta_k} \tanh^n r_k\,
|n^s_{\bm k}, n^s_{-\bm k}\rangle_{b}\,,
\end{equation}
where $|n^s_{\bm k}, n^s_{-\bm k}\rangle_{b} = \frac{1}{n!} (\hat{b}_{\bm k}^{s\dagger})^n (\hat{b}_{-\bm k}^{s\dagger})^n |0\rangle_{b}$.
This expression makes it explicit that cosmological particle production
creates graviton pairs with momenta $\bm k$ and $-\bm k$, namely a
two-mode squeezed vacuum state.

We next consider a simple cosmological model in which an inflationary
(de Sitter) phase is followed instantaneously by a radiation-dominated
phase. We assume the transition occur at conformal time $\eta=\eta_1>0$. A convenient choice of scale factor is
\begin{equation}
a(\eta)=
\begin{cases}
-\dfrac{1}{H(\eta-2\eta_1)},
& -\infty<\eta<\eta_1,
\\[1.2ex]
\dfrac{\eta}{H\eta_1^2},
& \eta_1<\eta<\eta_{\rm eq},
\end{cases}
\label{eq:scale-factor-piecewise}
\end{equation}
where $H$ is the Hubble parameter during inflation and $\eta_{\rm eq}$ is the conformal time at matter-radiation equality. 
In this standard inflation model, the equation of motion \eqref{eq:eom} yields the mode functions for the inflationary and radiation phases, respectively, as
\begin{align}
f_{k}(\eta)\equiv\frac{1}{\sqrt{2k}}\left(1-\frac{i}{k \left( \eta -2\eta_1 \right)}\right)e^{-ik \left( \eta -2\eta_1 \right)}, \quad 
g_{k}(\eta)\equiv\frac{1}{\sqrt{2k}}\,e^{-ik \eta } \,.
\end{align}

By matching the mode functions and their first derivatives at $\eta=\eta_1$, the Bogoliubov coefficients are
determined as
\begin{align}\label{eq:Bogoliubov-coefficients}
\alpha_k =\left(1-\frac{1}{2k^2\eta_1^2}+\frac{i}{k\eta_1}\right) e^{2ik\eta_1},\quad 
\beta_k = -\frac{1}{2k^2\eta_1^2}\,,
\end{align}
and the squeezing parameter $r_k$ reads, 
\begin{equation}\label{eq:squeezing-parameter-inflation}
\sinh r_k=\biggl|\frac{1}{2k^2\eta_1^2}\biggr|\,.
\end{equation}
Therefore, the long-wavelength modes satisfying $k\eta_1\ll 1$ become highly squeezed, whereas the short-wavelength modes with $k\eta_1\gg 1$ remain close to the adiabatic vacuum.

Next, we compute the equal-time two-point function of the primordial gravitons.
Since the physical initial condition is specified by the Bunch-Davies vacuum state $|0\rangle_a$,
the relevant correlator is
\begin{equation}
\label{eq:late-time-correlator-def}
\langle \hat h_{ij}(\eta,\bm x) \hat h_{lm}(\eta,\bm y)\rangle_a \equiv
{}_a\!\bra{0}\,
\hat h_{ij}(\eta,\bm x)\hat h_{lm}(\eta,\bm y)\,
\ket{0}_a ,
\end{equation}
where $\eta>\eta_1$ is taken in the radiation-dominated phase.

By using the mode expansion of the canonically normalized field 
$\hat{\tilde h}_{\bm k}^s(\eta)$ in the radiation phase,
\begin{equation}
\hat{\tilde h}_{\bm k}^s(\eta)
= g_k(\eta)\hat b_{\bm k}^s
+ g_k^*(\eta)\hat b_{-\bm k}^{s\dagger}\,, \quad
g_k(\eta)=\frac{1}{\sqrt{2k}}e^{-ik\eta}\,,
\end{equation}
with the Bogoliubov transformation~\eqref{eq:Bogoliubov-transformation}, the operator $\hat{\tilde h}_{\bm k}^s(\eta)$ can be rewritten in terms of the initial creation and annihilation operators as
\begin{align}
\hat{\tilde h}_{\bm k}^s(\eta)
&=
\Bigl[\alpha_k^* g_k(\eta)-\beta_k g_k^*(\eta)\Bigr]\hat a_{\bm k}^s
+
\Bigl[\alpha_k g_k^*(\eta)-\beta_k^* g_k(\eta)\Bigr]\hat a_{-\bm k}^{s\dagger}
\nonumber\\
&=
f_k(\eta)\hat a_{\bm k}^s
+
f_k^*(\eta)\hat a_{-\bm k}^{s\dagger}\,.
\label{eq:fk-late-time-definition}
\end{align}
Thus, the initial mode function in the inflationary phase is written as
\begin{equation}
\label{eq:fk-explicit-late}
f_k(\eta)
=
\alpha_k^* g_k(\eta)-\beta_k g_k^*(\eta)
=
\frac{1}{\sqrt{2k}}
\left(
\alpha_k^* e^{-ik\eta}
-
\beta_k e^{ik\eta}
\right)\,.
\end{equation}
Since $\hat a_{\bm k}^s\ket{0}_a=0$, 
the two-point function of the canonically normalized field is given as
\begin{align}
{}_a\!\bra{0}\,
\hat{\tilde h}_{\bm k}^s(\eta)\hat{\tilde h}_{\bm p}^{s'}(\eta)
\,\ket{0}_a
&=
{}_a\!\bra{0}
\Bigl(f_k \hat a_{\bm k}^s + f_k^* \hat a_{-\bm k}^{s\dagger}\Bigr)
\Bigl(f_p \hat a_{\bm p}^{s'} + f_p^* \hat a_{-\bm p}^{s'\dagger}\Bigr)
\ket{0}_a
\nonumber\\
&=
f_k(\eta)f_p^*(\eta)\,
{}_a\!\bra{0}\hat a_{\bm k}^s \hat a_{-\bm p}^{s'\dagger}\ket{0}_a
\nonumber\\
&=
(2\pi)^3\delta^{ss'}\delta^{(3)}(\bm k+\bm p)\,
|f_k(\eta)|^2.
\label{eq:tildeh-correlator-momentum}
\end{align}

Therefore, we obtain
\begin{equation}
\label{eq:hk-correlator-momentum}
{}_a\!\bra{0}\,
\hat h_{\bm k}^s(\eta)\hat h_{\bm p}^{s'}(\eta)
\,\ket{0}_a
=(2\pi)^3\delta^{ss'}\delta^{(3)}(\bm k+\bm p)\,
\frac{|f_k(\eta)|^2}{a^2(\eta)}\,,
\end{equation}
where the numerator is given by
\begin{align}
|f_k(\eta)|^2
&= \frac{1}{2k}
\left( |\alpha_k|^2+|\beta_k|^2
- \alpha_k\beta_k^*e^{-2ik\eta}
- \alpha_k^*\beta_k e^{2ik\eta}
\right)
\nonumber\\
&= \frac{1}{2k}
\left[
1+2|\beta_k|^2
- 2\,{\rm Re}\!\left(\alpha_k\beta_k^*e^{-2ik\eta}\right)
\right].
\label{eq:fk-squared-general}
\end{align}
Substituting the Bogoliubov coefficients~\eqref{eq:Bogoliubov-coefficients} and the radiation-phase scale factor into the above, we obtain the following expression,
\begin{equation}
\label{eq:hk-squared-explicit}
\frac{|f_k(\eta)|^2}{a^2(\eta)}
= \frac{H^2\eta_1^4}{2k\eta^2}
\left[
1+\frac{1}{2k^4\eta_1^4}
+\left( \frac{1}{k^2\eta_1^2} -\frac{1}{2k^4\eta_1^4} \right)\cos \left(2k(\eta-\eta_1)\right)
+\frac{1}{k^3\eta_1^3}\sin\left(2k(\eta-\eta_1)\right)
\right]\,.
\end{equation}
From Eq.~\eqref{eq:fourier-tensor},  
the graviton two-point function is obtained as
\begin{align}
\label{eq:hijhlm-general}
\langle \hat h_{ij}(\eta,\bm x) \hat h_{lm}(\eta,\bm y)\rangle_a
&= \frac{2}{M_{\rm pl}^2}
\sum_s \int\frac{\mathrm{d}^3k}{(2\pi)^3}\,
e^{i\bm k\cdot(\bm x-\bm y)}
\,p_{ij}^s(\bm k)\,p_{lm}^{s*}(\bm k)\,
\frac{|f_k(\eta)|^2}{a^2(\eta)}
\nonumber\\
& = \frac{2}{M_{\rm pl}^2}
\int\frac{\mathrm{d}^3k}{(2\pi)^3}\,
e^{i\bm k\cdot(\bm x-\bm y)}
\,\Pi_{ijlm}(\bar{\bm k})\,
\frac{|f_k(\eta)|^2}{a^2(\eta)},
\end{align}
where
\begin{equation}
\Pi_{ijlm}(\bar{\bm k})
\equiv
\sum_s p_{ij}^s(\bm k)p_{lm}^{s*}(\bm k)
=
P_{il}P_{jm}+P_{im}P_{jl}-P_{ij}P_{lm},
\quad
P_{ij}\equiv \delta_{ij}-\bar k_i \bar k_j,
\end{equation}
is the transverse-traceless projector with the unit vector $\bar k_i=\frac{k_i}{|\bm k|}$.

It is instructive to examine the long-wavelength limit corresponding to $k\eta_1 \ll 1$ and $k\eta \ll 1$. In this limit, we expand 
\begin{align}
\cos\left(2k(\eta-\eta_1)\right) & = 1 - 2k^2(\eta-\eta_1)^2 + \mathcal{O}(k^4)\,, \\
\sin\left(2k(\eta-\eta_1)\right) & = 2k(\eta-\eta_1) + \mathcal{O}(k^3)\,,
\end{align}
and find that the bracket in Eq.~\eqref{eq:hk-squared-explicit} behaves as
\begin{equation}
1+\frac{1}{2k^4\eta_1^4}
+\left( \frac{1}{k^2\eta_1^2} -\frac{1}{2k^4\eta_1^4} \right)\cos \left(2k(\eta-\eta_1)\right)
+\frac{1}{k^3\eta_1^3}\sin\left(2k(\eta-\eta_1)\right)
=
\frac{\eta^2}{k^2\eta_1^4}
+\mathcal{O}(k^0).
\end{equation}
Thus, we recover the standard result on superhorizon scales,
\begin{equation}
\label{eq:superhorizon-freezeout}
\frac{|f_k(\eta)|^2}{a^2(\eta)} \simeq \frac{H^2}{2k^3}.
\end{equation}
Substituting this into Eq.~\eqref{eq:hijhlm-general}, we obtain
\begin{equation}
\label{eq:hijhlm-superhorizon}
\langle \hat h_{ij}(\eta,\bm x) \hat h_{lm}(\eta,\bm y)\rangle_a
\simeq
\frac{H^2}{M_{\rm pl}^2}
\int\frac{\mathrm{d}^3k}{(2\pi)^3}\,
\frac{e^{i\bm k\cdot(\bm x-\bm y)}}{k^3}\,
\Pi_{ijlm}(\bar{\bm k}).
\end{equation}
The corresponding power spectrum is
\begin{equation}
\langle \hat h_{ij}(\eta,\bm x) \hat h_{lm}(\eta,\bm x)\rangle_a
\equiv \int \mathcal{P}_T(k) \frac{\mathrm{d}k}{k},\quad 
\mathcal{P}_T(k)
=
\frac{4k^3}{\pi^2 M_{\rm pl}^2}\,
\frac{|f_k(\eta)|^2}{a^2(\eta)},
\end{equation}
which reduces in the superhorizon limit to
\begin{equation}
\mathcal{P}_T(k)
\simeq
\frac{2H^2}{\pi^2 M_{\rm pl}^2}.
\end{equation}

Next, we evaluate the graviton two-point correlation function by introducing appropriate cutoffs. Since the superhorizon approximation~\eqref{eq:superhorizon-freezeout} is valid only for modes satisfying $k\eta\ll1$, we restrict the integration with an \ac{ir} cutoff $k_{\rm IR}$ and a \ac{uv} cutoff $k_{\rm UV}$~\cite{Giddings:2010nc,Giddings:2011zd},
\begin{equation}
\label{eq:hijhlm-superhorizon-cutoff}
\langle \hat h_{ij}(\eta,\bm x)\hat h_{lm}(\eta,\bm x)\rangle_a
\simeq
\frac{2H^2}{\pi^2 M_{\rm pl}^2}
\int_{k_{\rm IR}}^{k_{\rm UV}}\frac{\mathrm{d}k}{k}
=\frac{2H^2}{\pi^2 M_{\rm pl}^2}
\ln\!\left(\frac{k_{\rm UV}}{k_{\rm IR}}\right).
\end{equation}
Parameterizing the hierarchy of these cutoffs by the effective number of inflationary e-folds,
\begin{equation}
N_{\rm eff}\equiv \ln\!\left(\frac{k_{\rm UV}}{k_{\rm IR}}\right),
\end{equation}
we obtain
\begin{equation}\label{eq:inflation_graviton_flctuations}
\langle \hat h^2 \rangle_a
\sim
\frac{2H^2}{\pi^2 M_{\rm pl}^2}\,N_{\rm eff}.
\end{equation}

This estimate captures the long-wavelength \ac{ir} contribution but omits the short-distance \ac{uv} part expected from standard \ac{qft}. Indeed, in the high-$k$ limit the full mode-function expression~\eqref{eq:hk-squared-explicit} reduces to
\begin{equation}
\frac{|f_k(\eta)|^2}{a^2(\eta)}
\simeq \frac{H^2\eta_1^4}{2\eta^2}\,\frac{1}{k},
\end{equation}
so that the graviton two-point function acquires a \ac{uv} contribution
\begin{equation}
\langle \hat h_{ij}(\eta,\bm x)\hat h_{lm}(\eta,\bm x)\rangle_a
\sim \frac{H^2\eta_1^4}{M_{\rm pl}^2\eta^2}
\int^{\Lambda_{\rm UV}}\! \mathrm{d}k\, k,
\end{equation}
which is quadratically divergent. This divergence is not specific to primordial gravitons but is the standard short-distance divergence of a local \ac{qft}. An analogous quadratic divergence appears for graviton fluctuations in Minkowski spacetime,
\begin{equation}\label{eq:Minkowski_graviton_flctuations}
\langle \hat h^2 \rangle
\sim
\frac{\Lambda_{\rm UV}^2}{M_{\rm pl}^2}.
\end{equation}

A systematic treatment of this \ac{uv} divergence would require a proper renormalization prescription. In the present work, we instead regulate the integral with an explicit cutoff and distinguish the inflationary long-wavelength contribution from the \ac{uv} contribution in the primordial graviton two-point function. This distinction is essential for our purposes, since the light-cone smearing studied in this work depends on how these two contributions are defined and regulated. Although the separation is not unique, the cutoff-based treatment adopted here provides a phenomenological estimate of the effect of primordial gravitons on light-cone fluctuations.

\section{Retarded Green's function in flat spacetime}
\label{app:QFT-basics}

In this appendix, we present the basic derivation of the retarded Green's function for a free scalar field in flat spacetime used in Sec.~\ref{sec:retarded-Greens-function}. In the framework of \ac{qft}, the causal propagation of field disturbances is encoded in the commutator of field operators. We consider a free real scalar field with the mode expansion
\begin{equation}
\hat{\phi}(x) = \int \frac{d^3k}{(2\pi)^3\,2\omega_k}\left(\hat{a}_{\mathbf{k}}\,e^{ik\cdot x} + \hat{a}_{\mathbf{k}}^\dagger\,e^{-ik\cdot x}\right),
\end{equation}
where $k\cdot x = -\omega_k\,t + \mathbf{k}\cdot\mathbf{x}$, and the creation and annihilation operators obey the relativistically normalized commutation relation
\begin{equation}
[\hat{a}_{\mathbf{k}}, \hat{a}_{\mathbf{k}'}^\dagger] = (2\pi)^3\,2\omega_k\,\delta^{(3)}(\mathbf{k}-\mathbf{k}').
\end{equation}

For a free field, $[\hat\phi(x),\hat\phi(x')]$ is a $c$-number, and we define the Pauli-Jordan function $\Delta(x-x')$ by
\begin{equation}
[\hat{\phi}(x),\hat{\phi}(x')] = i\Delta(x-x') = \langle 0|[\hat{\phi}(x),\hat{\phi}(x')]|0\rangle.
\end{equation}
Inserting the mode expansion and using the canonical commutation relation, direct calculation gives
\begin{align}
\begin{split}
\langle 0|[\hat{\phi}(x),\hat{\phi}(x')]|0\rangle
&= \int\frac{d^3k}{(2\pi)^3\,2\omega_k}\left(e^{ik\cdot(x-x')}-e^{-ik\cdot(x-x')}\right)
\\
&= -i\int\frac{d^3k}{(2\pi)^3\,\omega_k}\,\sin\!\bigl[\omega_k(t-t')-\mathbf{k}\cdot(\mathbf{x}-\mathbf{x}')\bigr]
\\
&= -i\int\frac{d^3k}{(2\pi)^3\,\omega_k}\,\sin\!\bigl[\omega_k(t-t')\bigr]\,e^{i\mathbf{k}\cdot(\mathbf{x}-\mathbf{x}')},
\end{split}
\end{align}
where in the last step the term odd under $\mathbf{k}\to-\mathbf{k}$ has been dropped.

The retarded Green's function is defined in terms of this commutator by
\begin{equation}\label{eq:Gret-from-commutator}
G_{\rm ret}(x,x') \equiv i\,\theta(t-t')\,\langle 0|[\hat\phi(x),\hat\phi(x')]|0\rangle,
\end{equation}
and is the causal fundamental solution of the wave equation
\begin{equation}\label{eq:Gret-wave-eq}
\Box\,G_{\rm ret}(x,x') = -\delta^{(4)}(x-x'),
\quad
G_{\rm ret}(x,x') = 0 \ \ \text{for}\ \ t<t',
\end{equation}
where $\Box = \eta^{\mu\nu}\partial_\mu\partial_\nu = -\partial_t^2 + \nabla^2$ is the d'Alembertian in our signature. Specializing to the massless case $\omega_k=|\mathbf{k}|$, we obtain
\begin{align}
\begin{split}
G_{\rm ret}^{(0)}(x,x')
&= \theta(t-t')\int\frac{d^3k}{(2\pi)^3}\,\frac{\sin\!\bigl[\omega_k(t-t')\bigr]}{\omega_k}\,e^{i\mathbf{k}\cdot(\mathbf{x}-\mathbf{x}')}
\\
&= \frac{\theta(t-t')\,\delta\left((t-t')-|\mathbf{x}-\mathbf{x}'|\right)}{4\pi\,|\mathbf{x}-\mathbf{x}'|}
\\
&= \frac{\theta(t-t')}{4\pi}\,\delta(\sigma_0),
\end{split}
\end{align}
where $\sigma_0 = \frac{1}{2}\eta_{\mu\nu}(x-x')^\mu(x-x')^\nu$ is the unperturbed Synge's world function. The retarded Green's function thus has support strictly on the future light cone ($\sigma_0=0$, $t>t'$), and vanishes elsewhere.

\section{Evaluation of smeared Feynman propagator}
\label{app:smeared-feynman-propagator}
In this appendix, we provide the smeared Feynman propagator appearing in the main text~\cite{Ford:1994cr}.
For a massless scalar field, the Feynman propagator can be written as
\begin{equation}
G_F(x,x') = -\frac{1}{8\pi^2}
\left[\frac{i}{\sigma}+\pi\delta(\sigma)\right].
    \label{eq:Feynman-distribution}
\end{equation}
Equivalently, using
\begin{equation}
\int_0^\infty du\,e^{iu\sigma}
= \pi\delta(\sigma) +i\,\mathcal{P}\frac{1}{\sigma},
\end{equation}
where $\mathcal{P}$ denotes the Cauchy principal value, 
we can write
\begin{equation}
G_F(x,x')=
-\frac{1}{8\pi^2}\int_0^\infty du\,e^{iu\sigma}.
\label{eq:Feynman-integral-representation}
\end{equation}

Substituting $\sigma=\sigma_0+\hat{\sigma}_1$ and averaging over the graviton fluctuations, we find
\begin{equation}
    \left\langle G_F(x,x')\right\rangle
    =-\frac{1}{8\pi^2}
    \int_0^\infty du\,
    e^{iu\sigma_0}
    e^{-\frac{1}{2}u^2\langle \hat{\sigma}_1^2 \rangle}\,.
    \label{eq:smeared-Feynman-integral}
\end{equation}
Separating real and imaginary parts gives
\begin{align}
\left\langle G_F(x,x')\right\rangle
& = -\frac{1}{8\pi^2}
\int_0^\infty du\,
\cos(\sigma_0u)e^{-\frac{1}{2}u^2\langle \hat{\sigma}_1^2 \rangle}  -\frac{i}{8\pi^2}
\int_0^\infty du\,
\sin(\sigma_0u)e^{-\frac{1}{2}u^2\langle \hat{\sigma}_1^2 \rangle} 
\nonumber \\
& =-\frac{1}{8\pi^2}
    \sqrt{
        \frac{\pi}{2\langle \hat{\sigma}_1^2\rangle}
    }
    \exp\left[
        -\frac{\sigma_0^2}{2\langle \hat{\sigma}_1^2\rangle}
    \right] -\frac{i}{8\pi^2}
    \int_0^\infty du\,
    \sin(\sigma_0u)e^{-\frac{1}{2}u^2\langle \hat{\sigma}_1^2 \rangle}\,.
    \label{eq:smeared-Feynman-separated}
\end{align}
The first term is the smeared light-cone
$\delta$-function. In the limit $\langle \hat{\sigma}_1^2\rangle\to0^+$, we obtain 
\begin{equation}
\frac{1}{\sqrt{2\pi\langle \hat{\sigma}_1^2\rangle}}
\exp\left[ -\frac{\sigma_0^2}{2\langle \hat{\sigma}_1^2\rangle}
\right]\ 
\longrightarrow \ 
\delta(\sigma_0)\,.
\end{equation}
The second term is the smeared version of the principal-value singularity $1/\sigma_0$, and in the same limit,
\begin{equation}
\int_0^\infty du\,
\sin(\sigma_0u) \exp\left[
-\frac{1}{2}u^2\langle \hat{\sigma}_1^2\rangle \right]\
\longrightarrow \
\mathcal{P}\frac{1}{\sigma_0}.
\end{equation}
Therefore Eq.~\eqref{eq:smeared-Feynman-separated} reproduces the standard
massless Feynman propagator in the limit of vanishing light-cone fluctuations. At the classical light cone $\sigma_0=0$ and for a fixed nonzero point-separated variance, however, the smeared propagator is finite,
\begin{equation}
\left\langle G_F(x,x')\right\rangle \ 
\longrightarrow \ -\frac{1}{8\pi^2} \sqrt{ \frac{\pi}{2\langle \hat{\sigma}_1^2\rangle}}\,.
\end{equation}
Thus, the singularity on the light cone is replaced by a
Gaussian peak of finite width
$\sqrt{\langle \hat{\sigma}_1^2\rangle}$.

Next, we consider the square of the Feynman propagator, since one-loop
amplitudes often contain products of propagators. Starting from
Eq.~\eqref{eq:Feynman-integral-representation}, we obtain
\begin{equation}
G_F^2(x,x') =
\frac{1}{(8\pi^2)^2}\int_0^\infty du \int_0^\infty dv\,
e^{i(u+v)\sigma}\,.
\end{equation}
After substituting $\sigma=\sigma_0+\hat{\sigma}_1$ and averaging over the graviton fluctuations, this becomes
\begin{equation}
\left\langle G_F^2(x,x')\right\rangle =
\frac{1}{(8\pi^2)^2}\int_0^\infty du\int_0^\infty dv\,
e^{i(u+v)\sigma_0}e^{-\frac{1}{2}(u+v)^2 \langle \hat{\sigma}_1^2\rangle}\,.
\end{equation}
Introducing the variables
\begin{equation}
s=u+v, \quad r=\frac{u}{u+v},
\end{equation}
with $0\leq s<\infty$, $0\leq r\leq1$, and Jacobian
$dudv=sdsdr$, we obtain
\begin{equation}
\left\langle G_F(z)^2\right\rangle
=
\frac{1}{64\pi^4}
\int_0^\infty ds\,
s\,
\exp\left[
is\sigma_0
-\frac{1}{2}s^2
\langle \hat{\sigma}_1^2\rangle
\right].
\label{eq:smeared_GF_square_compact}
\end{equation}

Separating real and imaginary parts gives
\begin{align}
    \left\langle G_F^2(x,x')\right\rangle
    =
    &\frac{1}{64\pi^4}
    \int_0^\infty ds\,
    s\cos(\sigma_0s)
    \exp\left[
        -\frac{1}{2}s^2
        \langle \hat{\sigma}_1^2\rangle
    \right]
    \nonumber \\
    &+
    \frac{i}{64\pi^4}
    \int_0^\infty ds\,
    s\sin(\sigma_0s)
    \exp\left[
        -\frac{1}{2}s^2
        \langle \hat{\sigma}_1^2\rangle
    \right].
    \label{eq:smeared-GF-square-real-imag}
\end{align}
The imaginary part can be evaluated explicitly, 
\begin{equation}
\int_0^\infty ds\,
s\sin(\sigma_0s)
e^{-\frac{1}{2}\langle \hat{\sigma}_1^2\rangle s^2}
=\sqrt{\frac{\pi}{2}}\,
\frac{\sigma_0}
{\langle \hat{\sigma}_1^2\rangle^{3/2}}
\exp\left[ -\frac{\sigma_0^2}{2\langle \hat{\sigma}_1^2\rangle}
\right]\,.
\end{equation}
Therefore, we obtain 
\begin{align}
\left\langle G_F^2(x,x')\right\rangle =
&\frac{1}{64\pi^4} \int_0^\infty ds\,
 s\cos(\sigma_0s)
e^{-\frac{1}{2}s^2\langle \hat{\sigma}_1^2 \rangle} 
\nonumber \\
&+ i\,
\frac{\sqrt{2\pi}\,\sigma_0}
{128\pi^4
\langle \hat{\sigma}_1^2\rangle^{3/2}}
\exp\left[ -\frac{\sigma_0^2}{2\langle \hat{\sigma}_1^2\rangle}
\right].
\label{eq:smeared-GF-square}
\end{align}
In particular, on the classical light cone $\sigma_0=0$, 
the imaginary part vanishes and the real part becomes
\begin{equation}
\left\langle G_F^2(x,x')\right\rangle \ 
\longrightarrow \ 
\frac{1}{64\pi^4}\int_0^\infty ds\,s\,
e^{-\frac{1}{2}
\langle \hat{\sigma}_1^2\rangle s^2} =
\frac{1}
{64\pi^4\langle \hat{\sigma}_1^2\rangle}.
\end{equation}
Hence, for a fixed nonzero point-separated variance, the square of the propagator is also finite on the classical light cone.

\bibliography{Refs}

@article{Dyson:2013hbl,
    author = "Dyson, Freeman",
    title = "{Is a graviton detectable?}",
    doi = "10.1142/S0217751X1330041X",
    journal = "Int. J. Mod. Phys. A",
    volume = "28",
    pages = "1330041",
    year = "2013"
}

@article{Blencowe:2012mp,
    author = "Blencowe, M. P.",
    title = "{Effective Field Theory Approach to Gravitationally Induced Decoherence}",
    eprint = "1211.4751",
    archivePrefix = "arXiv",
    primaryClass = "quant-ph",
    doi = "10.1103/PhysRevLett.111.021302",
    journal = "Phys. Rev. Lett.",
    volume = "111",
    number = "2",
    pages = "021302",
    year = "2013"
}

@article{DeLorenci:2014vwa,
    author = "De Lorenci, V. A. and Ford, L. H.",
    title = "{Decoherence induced by long wavelength gravitons}",
    eprint = "1412.4685",
    archivePrefix = "arXiv",
    primaryClass = "gr-qc",
    doi = "10.1103/PhysRevD.91.044038",
    journal = "Phys. Rev. D",
    volume = "91",
    number = "4",
    pages = "044038",
    year = "2015"
}

@article{Oniga:2015lro,
    author = "Oniga, Teodora and Wang, Charles H. -T.",
    title = "{Quantum gravitational decoherence of light and matter}",
    eprint = "1511.06678",
    archivePrefix = "arXiv",
    primaryClass = "quant-ph",
    doi = "10.1103/PhysRevD.93.044027",
    journal = "Phys. Rev. D",
    volume = "93",
    pages = "044027",
    year = "2016"
}

@article{Bassi:2017szd,
    author = "Bassi, Angelo and Gro{\ss}ardt, Andr{\'e} and Ulbricht, Hendrik",
    title = "{Gravitational Decoherence}",
    eprint = "1706.05677",
    archivePrefix = "arXiv",
    primaryClass = "quant-ph",
    doi = "10.1088/1361-6382/aa864f",
    journal = "Class. Quant. Grav.",
    volume = "34",
    number = "19",
    pages = "193002",
    year = "2017"
}

@article{Lagouvardos:2020laf,
    author = "Lagouvardos, Michalis and Anastopoulos, Charis",
    title = "{Gravitational decoherence of photons}",
    eprint = "2011.08270",
    archivePrefix = "arXiv",
    primaryClass = "gr-qc",
    doi = "10.1088/1361-6382/abf2f3",
    journal = "Class. Quant. Grav.",
    volume = "38",
    number = "11",
    pages = "115012",
    year = "2021"
}

@article{Guerreiro:2019vbq,
    author = "Guerreiro, Thiago",
    title = "{Quantum Effects in Gravity Waves}",
    eprint = "1911.11593",
    archivePrefix = "arXiv",
    primaryClass = "quant-ph",
    doi = "10.1088/1361-6382/ab9d5d",
    journal = "Class. Quant. Grav.",
    volume = "37",
    number = "15",
    pages = "155001",
    year = "2020"
}

@article{Parikh:2020nrd,
    author = "Parikh, Maulik and Wilczek, Frank and Zahariade, George",
    title = "{The Noise of Gravitons}",
    eprint = "2005.07211",
    archivePrefix = "arXiv",
    primaryClass = "hep-th",
    doi = "10.1142/S0218271820420018",
    journal = "Int. J. Mod. Phys. D",
    volume = "29",
    number = "14",
    pages = "2042001",
    year = "2020"
}

@article{Kanno:2020usf,
    author = "Kanno, Sugumi and Soda, Jiro and Tokuda, Junsei",
    title = "{Noise and decoherence induced by gravitons}",
    eprint = "2007.09838",
    archivePrefix = "arXiv",
    primaryClass = "hep-th",
    reportNumber = "OU-HET-1065, KOBE-COSMO-20-12",
    doi = "10.1103/PhysRevD.103.044017",
    journal = "Phys. Rev. D",
    volume = "103",
    number = "4",
    pages = "044017",
    year = "2021"
}

@article{Parikh:2020kfh,
    author = "Parikh, Maulik and Wilczek, Frank and Zahariade, George",
    title = "{Quantum Mechanics of Gravitational Waves}",
    eprint = "2010.08205",
    archivePrefix = "arXiv",
    primaryClass = "hep-th",
    doi = "10.1103/PhysRevLett.127.081602",
    journal = "Phys. Rev. Lett.",
    volume = "127",
    number = "8",
    pages = "081602",
    year = "2021"
}

@article{Parikh:2020fhy,
    author = "Parikh, Maulik and Wilczek, Frank and Zahariade, George",
    title = "{Signatures of the quantization of gravity at gravitational wave detectors}",
    eprint = "2010.08208",
    archivePrefix = "arXiv",
    primaryClass = "hep-th",
    doi = "10.1103/PhysRevD.104.046021",
    journal = "Phys. Rev. D",
    volume = "104",
    number = "4",
    pages = "046021",
    year = "2021"
}

@article{Kanno:2021gpt,
    author = "Kanno, Sugumi and Soda, Jiro and Tokuda, Junsei",
    title = "{Indirect detection of gravitons through quantum entanglement}",
    eprint = "2103.17053",
    archivePrefix = "arXiv",
    primaryClass = "gr-qc",
    reportNumber = "KOBE-COSMO-21-06",
    doi = "10.1103/PhysRevD.104.083516",
    journal = "Phys. Rev. D",
    volume = "104",
    number = "8",
    pages = "083516",
    year = "2021"
}

@article{Tobar:2023ksi,
    author = "Tobar, Germain and Manikandan, Sreenath K. and Beitel, Thomas and Pikovski, Igor",
    title = "{Detecting single gravitons with quantum sensing}",
    eprint = "2308.15440",
    archivePrefix = "arXiv",
    primaryClass = "quant-ph",
    reportNumber = "NORDITA 2023-040",
    doi = "10.1038/s41467-024-51420-8",
    journal = "Nature Commun.",
    volume = "15",
    number = "1",
    pages = "7229",
    year = "2024"
}

@article{Hsiang:2024qou,
    author = "Hsiang, Jen-Tsung and Cho, Hing-Tong and Hu, Bei-Lok",
    title = "{Graviton Physics: A Concise Tutorial on the Quantum Field Theory of Gravitons, Graviton Noise, and Gravitational Decoherence}",
    eprint = "2405.11790",
    archivePrefix = "arXiv",
    primaryClass = "hep-th",
    doi = "10.3390/universe10080306",
    journal = "Universe",
    volume = "10",
    number = "8",
    pages = "306",
    year = "2024"
}

@article{Carney:2023nzz,
    author = "Carney, Daniel and Domcke, Valerie and Rodd, Nicholas L.",
    title = "{Graviton detection and the quantization of gravity}",
    eprint = "2308.12988",
    archivePrefix = "arXiv",
    primaryClass = "hep-th",
    reportNumber = "CERN-TH-2023-155",
    doi = "10.1103/PhysRevD.109.044009",
    journal = "Phys. Rev. D",
    volume = "109",
    number = "4",
    pages = "044009",
    year = "2024"
}

@article{Kanno:2018cuk,
    author = "Kanno, Sugumi and Soda, Jiro",
    title = "{Detecting nonclassical primordial gravitational waves with Hanbury-Brown{\textendash}Twiss interferometry}",
    eprint = "1810.07604",
    archivePrefix = "arXiv",
    primaryClass = "hep-th",
    reportNumber = "OU-HET-980, KOBE-COSMO-18-09",
    doi = "10.1103/PhysRevD.99.084010",
    journal = "Phys. Rev. D",
    volume = "99",
    number = "8",
    pages = "084010",
    year = "2019"
}

@article{Kanno:2019gqw,
    author = "Kanno, Sugumi",
    title = "{Nonclassical primordial gravitational waves from the initial entangled state}",
    eprint = "1905.06800",
    archivePrefix = "arXiv",
    primaryClass = "hep-th",
    reportNumber = "OU-HET-1017",
    doi = "10.1103/PhysRevD.100.123536",
    journal = "Phys. Rev. D",
    volume = "100",
    number = "12",
    pages = "123536",
    year = "2019"
}

@article{Kanno:2024gjt,
    author = "Kanno, Sugumi and Matsui, Hiroki and Mukohyama, Shinji",
    title = "{Hanbury-Brown-Twiss interferometry and quantum nature of primordial gravitational waves in Ho{\v{r}}ava-Lifshitz gravity}",
    eprint = "2412.19514",
    archivePrefix = "arXiv",
    primaryClass = "gr-qc",
    reportNumber = "YITP-24-160, IPMU24-0044",
    doi = "10.1103/PhysRevD.111.104077",
    journal = "Phys. Rev. D",
    volume = "111",
    number = "10",
    pages = "104077",
    year = "2025"
}

@article{Kanno:2025how,
    author = "Kanno, Sugumi and Soda, Jiro and Taniguchi, Akira",
    title = "{Coherent State Description of Gravitational Waves from Binary Black Holes}",
    eprint = "2508.17947",
    archivePrefix = "arXiv",
    primaryClass = "gr-qc",
    reportNumber = "YITP-25-129, KOBE-COSMO-25-16",
    doi = "10.1103/kv1t-j27m",
    journal = "Phys. Rev. Lett.",
    volume = "136",
    number = "6",
    pages = "061404",
    year = "2026"
}

@article{Kanno:2025fpz,
    author = "Kanno, Sugumi and Soda, Jiro and Taniguchi, Akira",
    title = "{Binary gravitational waves as probes of quantum graviton states}",
    eprint = "2510.23326",
    archivePrefix = "arXiv",
    primaryClass = "gr-qc",
    reportNumber = "YITP-25-169, KOBE-COSMO-25-17",
    month = "10",
    year = "2025"
}

@article{Deser:1957zz,
    author = "Deser, Stanley",
    title = "{General Relativity and the Divergence Problem in Quantum Field Theory}",
    doi = "10.1103/RevModPhys.29.417",
    journal = "Rev. Mod. Phys.",
    volume = "29",
    pages = "417",
    year = "1957"
}

@article{Isham:1970aw,
    author = "Isham, C. J. and Salam, Abdus and Strathdee, J. A.",
    title = "{Infinity suppression gravity modified quantum electrodynamics}",
    reportNumber = "IC-70-131",
    doi = "10.1103/PhysRevD.3.1805",
    journal = "Phys. Rev. D",
    volume = "3",
    pages = "1805--1817",
    year = "1971"
}

@article{Isham:1972pf,
    author = "Isham, C. J. and Salam, Abdus and Strathdee, J. A.",
    title = "{Infinity suppression in gravity modified electrodynamics. II}",
    reportNumber = "IC-73-14",
    doi = "10.1103/PhysRevD.5.2548",
    journal = "Phys. Rev. D",
    volume = "5",
    pages = "2548--2565",
    year = "1972"
}

@article{Ford:1994cr,
    author = "Ford, L. H.",
    title = "{Gravitons and light cone fluctuations}",
    eprint = "gr-qc/9410047",
    archivePrefix = "arXiv",
    reportNumber = "TUTP-94-15",
    doi = "10.1103/PhysRevD.51.1692",
    journal = "Phys. Rev. D",
    volume = "51",
    pages = "1692--1700",
    year = "1995"
}

@article{Garay:1994en,
    author = "Garay, Luis J.",
    title = "{Quantum gravity and minimum length}",
    eprint = "gr-qc/9403008",
    archivePrefix = "arXiv",
    reportNumber = "IMPERIAL-TP-93-94-20",
    doi = "10.1142/S0217751X95000085",
    journal = "Int. J. Mod. Phys. A",
    volume = "10",
    pages = "145--166",
    year = "1995"
}

@article{Grishchuk:1990bj,
    author = "Grishchuk, L. P. and Sidorov, Yu. V.",
    title = "{Squeezed quantum states of relic gravitons and primordial density fluctuations}",
    doi = "10.1103/PhysRevD.42.3413",
    journal = "Phys. Rev. D",
    volume = "42",
    pages = "3413--3421",
    year = "1990"
}

@article{Albrecht:1992kf,
    author = "Albrecht, Andreas and Ferreira, Pedro and Joyce, Michael and Prokopec, Tomislav",
    title = "{Inflation and squeezed quantum states}",
    eprint = "astro-ph/9303001",
    archivePrefix = "arXiv",
    reportNumber = "IMPERIAL-TP-92-93-21",
    doi = "10.1103/PhysRevD.50.4807",
    journal = "Phys. Rev. D",
    volume = "50",
    pages = "4807--4820",
    year = "1994"
}

@article{Polarski:1995jg,
    author = "Polarski, David and Starobinsky, Alexei A.",
    title = "{Semiclassicality and decoherence of cosmological perturbations}",
    eprint = "gr-qc/9504030",
    archivePrefix = "arXiv",
    reportNumber = "LMPM-95-4",
    doi = "10.1088/0264-9381/13/3/006",
    journal = "Class. Quant. Grav.",
    volume = "13",
    pages = "377--392",
    year = "1996"
}

@article{Maggiore:1999vm,
    author = "Maggiore, Michele",
    title = "{Gravitational wave experiments and early universe cosmology}",
    eprint = "gr-qc/9909001",
    archivePrefix = "arXiv",
    reportNumber = "IFUP-TH-20-99",
    doi = "10.1016/S0370-1573(99)00102-7",
    journal = "Phys. Rept.",
    volume = "331",
    pages = "283--367",
    year = "2000"
}

@article{Giddings:2010nc,
    author = "Giddings, Steven B. and Sloth, Martin S.",
    title = "{Semiclassical relations and IR effects in de Sitter and slow-roll space-times}",
    eprint = "1005.1056",
    archivePrefix = "arXiv",
    primaryClass = "hep-th",
    reportNumber = "CERN-PH-TH-2010-095",
    doi = "10.1088/1475-7516/2011/01/023",
    journal = "JCAP",
    volume = "01",
    pages = "023",
    year = "2011"
}

@article{Giddings:2011zd,
    author = "Giddings, Steven B. and Sloth, Martin S.",
    title = "{Cosmological observables, IR growth of fluctuations, and scale-dependent anisotropies}",
    eprint = "1104.0002",
    archivePrefix = "arXiv",
    primaryClass = "hep-th",
    reportNumber = "CERN-PH-TH-2011-070",
    doi = "10.1103/PhysRevD.84.063528",
    journal = "Phys. Rev. D",
    volume = "84",
    pages = "063528",
    year = "2011"
}

@article{Dorlis:2025zzz,
    author = "Dorlis, Panagiotis and Mavromatos, Nick E. and Sarkar, Sarben and Vlachos, Sotirios-Neilos",
    title = "{Superradiant Axionic Black-Hole Clouds as Seeds for Graviton Squeezing}",
    eprint = "2507.01689",
    archivePrefix = "arXiv",
    primaryClass = "gr-qc",
    reportNumber = "KCL-PH-TH/2025-15",
    doi = "10.1103/9crd-zj6l",
    journal = "Phys. Rev. Lett.",
    volume = "135",
    number = "15",
    pages = "151501",
    year = "2025"
}

@article{Takeda:2025cye,
    author = "Takeda, Hiroki and Tanaka, Takahiro",
    title = "{Quantum decoherence of gravitational waves}",
    eprint = "2502.18560",
    archivePrefix = "arXiv",
    primaryClass = "gr-qc",
    doi = "10.1103/PhysRevD.111.104080",
    journal = "Phys. Rev. D",
    volume = "111",
    number = "10",
    pages = "104080",
    year = "2025"
}

@article{Ohanian:1996ni,
    author = "Ohanian, Hans C.",
    title = "{Finite quantum electrodynamics with a gravitationally smeared propagator}",
    reportNumber = "PRINT-97-065 (RENSSELAER)",
    doi = "10.1103/PhysRevD.55.5140",
    journal = "Phys. Rev. D",
    volume = "55",
    pages = "5140--5146",
    year = "1997"
}

@article{Ohanian:1999fu,
    author = "Ohanian, Hans C.",
    title = "{Smearing of propagators by gravitational fluctuations on the Planck scale}",
    doi = "10.1103/PhysRevD.60.104051",
    journal = "Phys. Rev. D",
    volume = "60",
    pages = "104051",
    year = "1999"
}

@article{Wheeler:1955zz,
    author = "Wheeler, J. A.",
    title = "{Geons}",
    doi = "10.1103/PhysRev.97.511",
    journal = "Phys. Rev.",
    volume = "97",
    pages = "511--536",
    year = "1955"
}

@article{Wheeler:1957mu,
    author = "Wheeler, John A.",
    title = "{On the Nature of quantum geometrodynamics}",
    doi = "10.1016/0003-4916(57)90050-7",
    journal = "Annals Phys.",
    volume = "2",
    pages = "604--614",
    year = "1957"
}

@article{Modesto:2009qc,
    author = "Modesto, Leonardo and Nicolini, Piero",
    title = "{Spectral dimension of a quantum universe}",
    eprint = "0912.0220",
    archivePrefix = "arXiv",
    primaryClass = "hep-th",
    doi = "10.1103/PhysRevD.81.104040",
    journal = "Phys. Rev. D",
    volume = "81",
    pages = "104040",
    year = "2010"
}

@article{Padmanabhan:2020rba,
    author = "Padmanabhan, T.",
    title = "{Probing the Planck scale: The modification of the time evolution operator due to the quantum structure of spacetime}",
    eprint = "2006.06701",
    archivePrefix = "arXiv",
    primaryClass = "gr-qc",
    doi = "10.1007/JHEP11(2020)013",
    journal = "JHEP",
    volume = "11",
    pages = "013",
    year = "2020"
}

@article{Abel:2019ufz,
    author = "Abel, Steven and Dondi, Nicola Andrea",
    title = "{UV Completion on the Worldline}",
    eprint = "1905.04258",
    archivePrefix = "arXiv",
    primaryClass = "hep-th",
    reportNumber = "CP3-Origins-2019-19 DNRF90, IPPP/19/36",
    doi = "10.1007/JHEP07(2019)090",
    journal = "JHEP",
    volume = "07",
    pages = "090",
    year = "2019"
}

@article{Kan:2020vut,
    author = "Kan, Nahomi and Kuniyasu, Masashi and Shiraishi, Kiyoshi and Wu, Zhenyuan",
    title = "{Vacuum expectation values in nontrivial background space from three types of UV improved Green{\textquoteright}s functions}",
    eprint = "2004.07527",
    archivePrefix = "arXiv",
    primaryClass = "hep-th",
    doi = "10.1142/S0217751X21500019",
    journal = "Int. J. Mod. Phys. A",
    volume = "36",
    number = "01",
    pages = "2150001",
    year = "2021"
}

@article{synge1960special,
  title={The special theory},
  author={Synge, JL},
  journal={and idem, Relativity: The General Theory (Amsterdam: North-Holland, 1964), pp. 114ff},
  pages={36},
  year={1960}
}

@article{Ford:1996qc,
    author = "Ford, L. H. and Svaiter, N. F.",
    title = "{Gravitons and light cone fluctuations. 2: Correlation functions}",
    eprint = "gr-qc/9604052",
    archivePrefix = "arXiv",
    reportNumber = "TUTP-96-1",
    doi = "10.1103/PhysRevD.54.2640",
    journal = "Phys. Rev. D",
    volume = "54",
    pages = "2640--2646",
    year = "1996"
}
\bibliographystyle{JHEP}

\end{document}